\documentclass[a4paper,11pt]{article}
\pdfoutput=1
\usepackage{./auxi/jheppub}
\usepackage[utf8]{inputenc}

\usepackage{simplewick}
\usepackage{dsfont}
\usepackage{float}
\usepackage{pgfplots}
\pgfplotsset{compat=1.14}

\let\oldbfseries=\bfseries
\let\oldmdseries=\mdseries
\let\oldnormalfont=\normalfont
\renewcommand{\bfseries}{\oldbfseries\boldmath}
\renewcommand{\mdseries}{\oldmdseries\unboldmath}
\renewcommand{\normalfont}{\oldnormalfont\unboldmath}

\makeatletter
\newlength{\apb@width}
\newcommand{\autoparbox}[2][c]{\settowidth{\apb@width}{#2}\parbox[#1]{\apb@width}{#2}}

\makeatother

\def \ph{\phantom}
\DeclareMathOperator{\ad}{ad}
\DeclareMathOperator{\tr}{tr}

\def\Bm{{\mathcal{B}}}

\def\Em{{\mathcal{E}}}

\def\Gm{{\mathcal{G}}}
\def\Hm{{\mathcal{H}}}

\def\Nm{{\mathcal{N}}}
\def\Om{{\mathcal{O}}}

\def\Qb{{\bar{Q}}}
\def\Sb{{\bar{S}}}
\def\zb{{\bar{z}}}
\def\jbar{{\bar{\jmath}}}

\def\a{{\alpha}}

\def\ad{{\dot{\alpha}}}

\def\ph{\phantom}

\newcommand{\beq}{\begin{equation}}
\newcommand{\eeq}{\end{equation}}

\definecolor{nicegreen}{rgb}{0.1,0.6,0.1}

\newmuskip\pFqskip
\mathchardef\pFcomma=\mathcode`,

\makeatletter
\renewcommand*\env@matrix[1][\arraystretch]{%
  \edef\arraystretch{#1}%
  \hskip -\arraycolsep
  \let\@ifnextchar\new@ifnextchar
  \array{*\c@MaxMatrixCols c}}
\makeatother

\graphicspath{{./figures/}}

\title{\boldmath Bootstrapping Coulomb and Higgs branch operators}

\author{Aleix Gimenez-Grau}
\author{and Pedro Liendo}

\affiliation{DESY Hamburg, Theory Group, Notkestra{\ss}e 85, D-22607 Hamburg, Germany}

\emailAdd{aleix.gimenez@desy.de}
\emailAdd{pedro.liendo@desy.de}

\preprint{DESY 20-099}

\abstract{	We apply the numerical conformal bootstrap to correlators of Coulomb and Higgs branch operators in $4d$ $\mathcal{N}=2$ superconformal theories. We start by revisiting previous results on single correlators of Coulomb branch operators. 
In particular, we present improved bounds on OPE coefficients for some selected Argyres-Douglas models, and compare them to recent work where the same cofficients were obtained in the limit of large $r$ charge. There is solid agreement between all the approaches. The improved bounds can be used to extract an approximate spectrum of the Argyres-Douglas models, which can then be used as a guide in order to corner these theories to numerical islands in the space of conformal dimensions. When there is a flavor symmetry present, we complement the analysis by including mixed correlators of Coulomb branch operators and the moment map, a Higgs branch operator which sits in the same multiplet as the flavor current. After calculating the relevant superconformal blocks we apply the numerical machinery to the mixed system. We put general constraints on CFT data appearing in the new channels, with particular emphasis on the simplest Argyres-Douglas model with non-trivial flavor symmetry.
}

\begin{document} 
\maketitle
\flushbottom


\section{Introduction}
\label{sec:intro}

Four-dimensional $\Nm=2$ superconformal theories are interesting models that despite a significant amount of symmetry, show highly non-trivial dynamics and constitute a vast landscape of theories. In this work, we use modern conformal bootstrap techniques to study a canonical set of correlators that involve \textit{Coulomb} and \textit{Higgs} branch operators. These are operators that sit in short multiplets of the superconformal algebra and whose vevs parameterize the moduli space vacua.

Both Coulomb and Higgs branches are a common feature of $\Nm=2$ theories and have been used as a starting point for the ambitious program of classifying $\Nm=2$ SCFTs. Coulomb branch geometries are characterized by their complex dimension, known as the rank, and a classification of all possible rank-one scale-invariant geometries was recently proposed in a series of papers \cite{Argyres:2015ffa,Argyres:2015gha,Argyres:2016xua,Argyres:2016xmc} (for a review of later developments see \cite{Argyres:2020nrr}).
The Higgs branch is also a useful organizing principle, specially when taking into account its close connection to the VOAs associated to $4d$ $\Nm=2$ SCFTs \cite{Beem:2013sza}.
It was conjectured in \cite{Beem:2017ooy} that one can recover the Higgs branch by looking at the associated variety of the corresponding VOA. This observation lead to an elegant description of several VOAs in terms of free field realizations \cite{Bonetti:2018fqz,Beem:2019tfp,Beem:2019snk}, and constitutes a first step towards a classification of VOAs associated to $4d$ $\Nm=2$ theories. 

Progress in understanding $\Nm=2$ dynamics usually involves quantities protected by supersymmetry, and access to non-protected data is still a challenge. In this work, we will study correlators of Coulomb and Higgs branch operators using the numerical bootstrap of~\cite{Rattazzi:2008pe}.
Even though the external operators in our analysis are protected objects, they still capture an infinite amount of non-protected data in their correlators. Our results will be naturally split into two parts: the single-correlator and the mixed-correlator bootstrap. 

The single correlator bootstrap for each type of operator has already appeared in the literature \cite{Beem:2014zpa,Lemos:2015awa,Cornagliotto:2017snu}. 
Here we refine the Coulomb branch results motivated in part by the works~\cite{Hellerman:2018xpi,Grassi:2019txd}, where OPE coefficients of Coulomb branch operators for certain Argyres-Douglas models were calculated in the large $r$-charge limit.
The same OPE coefficients can be estimated by obtaining upper and lower bounds using the numerical bootstrap, and our improved results show solid agreement between the two approaches.

The improved OPE-coefficient bounds also give a better idea of the non-protected spectrum of these models. 
This is because numerical exclusion curves correspond to solutions to crossing with a spectrum that can be extracted from the zeros of the numerical functional \cite{ElShowk:2012hu}. We can then use the approximate spectrum as a guide and corner models by assuming gaps in the single correlator bootstrap, similarly to what was done in \cite{Li:2017kck}.

Having exhausted the single correlator analysis, we move on to the mixed-correlator bootstrap and include Higgs branch operators, which will give us access to a new region of parameter space that cannot be accessed by looking at single correlators separately. The mixed correlator setup requires the calculation of new superconformal blocks that include Coulomb and Higgs branch operators. We solve this problem by demanding that different $\Nm=1$ decompositions of the blocks are consistent, and contribute a new entry to the superblock catalog, a result interesting on its own right. With the mixed correlator at hand we explore the landscape of theories with $SU(2)$ flavor symmetry, and also attempt to corner an individual Argyres-Douglas model, whose symmetries are consistent with our setup. 

The rest of this work is organized as follows.
In section~\ref{sec:blocks} we review the properties of the Coulomb and Higgs branch operators that will be the main focus of this work.
We describe to which short representations of the superconformal algebra they belong, and introduce their OPE selection rules and superconformal blocks.
We also summarize the most salient features of the landscape of known $\Nm = 2$ superconformal field theories.
In section~\ref{sec:crossing} we present the crossing relations obeyed by our correlation functions. Special care is required for the four-point function of moment-map operators, where one needs to take into account the chiral algebra construction of~\cite{Beem:2013sza}.
In section~\ref{sec:numerics} we present the results obtained from applying numerical bootstrap techniques to the single and mixed correlators.
We conclude in section~\ref{sec:conclusions} with a summary of our results and an overview of possible future directions.
The derivation of the new superconformal blocks and the details of our numerical bootstrap setup are relegated to appendix~\ref{sec:app_blocks} and~\ref{sec:app_numerics}.


\section{Preliminaries}
\label{sec:blocks}

We start with a preliminary review of the two basic operators that will be the main focus of this work. We will concentrate on two types of short multiplets of the superconformal algebra, whose correlators capture ``canonical data'' of the theory, in the sense that it is data common to most $\Nm=2$ superconformal theories. In particular, we will study short operators whose vevs parameterize the Coulomb and Higgs branches of the moduli space of vacua.

\subsection{Canonical data in $\Nm=2$ SCFTs}

\subsubsection{Coulomb branch operators}

We denote Coulomb branch operators by $\varphi_r(x)$ and $\bar \varphi_{-r}(x)$. These are superconformal primaries killed by supercharges of the same Lorentz chirality, and their conformal dimension is fixed by supersymmetry in terms of their corresponding $r$ charges:
\begin{align}
	\left[\bar Q_{I\ad}, \, \varphi_r(0)   \right] = 0, \quad
	\left[Q^I_\a, \,   \bar \varphi_{-r}(0)\right] = 0 
	\qquad \Rightarrow \qquad
	\Delta_{\varphi_r} = \Delta_{\bar \varphi_{-r}} = \frac{1}{2} r.
\end{align}
In the notation of~\cite{Cordova:2016emh}, which we will use throughout the paper, $\varphi$ is the superprimary of a $ L \bar B [0;0]^{(0;r)}$ multiplet.\footnote{For readers familiar with the notation of~\cite{Dolan:2002zh} this corresponds to the $\Em_{r/2}$ multiplet.}
In the literature, they are sometimes called chiral primary operators, and we will often refer to them simply as \emph{chiral} or \emph{antichiral} operators.
They form a ring under the OPE, called the Coulomb branch chiral ring, and the number of generators in this ring defines the rank of the theory. Canonical data associated with this ring are the $U(1)_r$ charge values of the generators. In Lagrangian theories, the chiral ring generators are given by gauge invariant combinations $\tr \phi^n$ of the basic vector multiplet $\phi$, and their $r$-charges are always integer-valued. In interacting Lagrangian models, each coupling constant will have an associated chiral operator $\tr \phi^2$ of dimension $\Delta = 2$, which is the superconformal primary of the multiplet that contains the exactly marginal deformation responsible for the corresponding direction in the conformal manifold.

Non-lagrangian examples include Argyres-Douglas models, in which the $r$-charges of the Coulomb branch generators can have fractional values, and it is therefore unlikely they will have a standard Lagrangian description. One advantage of the bootstrap approach followed here is that the value of $r$ for chiral operators is a parameter in the crossing equations, and it can take any continuous value. This makes Argyres-Douglas models prime candidates for the conformal bootstrap.

Correlators involving chiral operators have been nevertheless studied by a variety of means, and some recent results in the literature will guide our analysis. One approach is the observation that certain chiral correlators satisfy exact differential equations known as the $tt^{*}$ equations~\cite{Baggio:2014ioa}. 
This lead to a series of exciting developments~\cite{Baggio:2014sna,Baggio:2015vxa,Baggio:2016skg,Gerchkovitz:2016gxx}, which provide an algorithmic prescription to calculate OPE coefficient among operators in the chiral ring of Lagrangian $\Nm = 2$ theories.
An alternative approach is provided by the study of these correlators in the limit of large $r$-charge~\cite{Hellerman:2015nra,Hellerman:2017sur,Beccaria:2018xxl,Bourget:2018obm}. 
Of particular relevance for us will be the works \cite{Hellerman:2018xpi,Grassi:2019txd}, from which one can extract OPE coefficients of Coulomb branch operators in rank-one Argyres-Douglas models.
We will be able to compare these results with our bootstrap bounds in section~\ref{sec:numerics}, showing consistent results between the two approaches.

\subsubsection{The moment map}

Complementary to the Coulomb branch there is also canonical data associated to the Higgs branch. This branch of moduli space is parameterized by another type of short multiplets killed by a different combination of supercharges, whose highest weights also form a ring under the OPE. In this work we will only consider the so-called moment map operator $M^A_{(IJ)}(x)$, which is the superconformal primary of the $ B \bar B[0;0]^{(2;0)}$ multiplet.\footnote{For readers familiar with the notation of~\cite{Dolan:2002zh} this corresponds to the $\hat \Bm_1$ multiplet.}
Unlike the previous case, this multiplet satisfies a shortening condition involving supercharges of both chiralities:
\begin{align}
\label{eq:short-M}
	\left[ \Qb_{(I}, \, M^A_{JK)}(0) \right] = 0, \quad
	\left[ Q_{(I}, \, M^A_{JK)}(0) \right] = 0 
	\qquad \Rightarrow \qquad
	\Delta_{M} = 2\, .
\end{align} 
This operator is neutral under $U(1)_r$, and transforms as a triplet under $SU(2)_R$, which we represent with two symmetric fundamental indices $I,J = 1, 2$.
What makes the moment map particularly important is that it belongs to the same multiplet as the flavor current $j_\mu^A$, 
this means that both transform in the adjoint of the flavor group $G_F$, which we indicate with an adjoint index $A = 1, \ldots, \dim G_F$. It follows from this discussion that the moment map will be present whenever there is a global (flavor) symmetry.
Flavor symmetries are ubiquitous in $\Nm=2$ superconformal theories, and correlators of flavor currents capture canonical data associated to the Higgs branch. Relevant for us is the flavor central charge $k$ which can be considered analogous to the most common central charge $c$, which is associated to correlators of the stress tensor.

A fact that will play a fundamental role in our subsequent analysis is that the moment map belongs to a special class of operators whose protected data is described by a $2d$ chiral algebra \cite{Beem:2013sza}.\footnote{
	Coulomb branch operators are not captured by the chiral algebra, and apart from protected conformal dimensions the rest of the CFT data appearing in their correlators is always dynamical.}
Because the moment map sits in the adjoint of the flavor group, the corresponding operator in the $2d$ chiral algebra is an affine Kac-Moody current. The dictionary between the $4d$ and $2d$ theories is well known:
\begin{align}
M^A_{(IJ)}(x) \to J^A(z)\, , \qquad k_{4d} \to -\frac{1}{2}k_{2d}\, .
\end{align} 
The chiral algebra description of protected data allows to solve for an infinite number of OPE coefficients in terms of the two central charges $c$ and $k$. 
This has two important consequences for us:
\begin{itemize}
	\item Imposing unitarity of the parent $4d$ theory requires that the calculated OPE coefficients are positive, leading to strict analytic unitarity bounds on the central charges $c$ and $k$~\cite{Beem:2013sza,Liendo:2015ofa,Lemos:2015orc,Beem:2018duj}. These unitarity bounds are a good organizing principle that we will use in the next section when we discuss the landscape of $\Nm = 2$ theories, in order to choose which theories one should focus on. 
	
	\item The second consequence is that having analytic control over the protected part of the correlator gives valuable input for the numerical bootstrap, which is mostly concerned with non-protected data. In the crossing equations we present below, the infinite number of short operators appearing in the moment map four-point function can be summed and treated exactly.
\end{itemize}

\subsubsection{Case studies}
\label{ref:case-studies}

Although the bootstrap is an agnostic approach to study SCFTs, it is important to keep in mind what assumptions we are making and what theories our bounds apply to.
For example, chiral operators will be present when the theory has a Coulomb branch of rank one or higher, and the moment map signals a global flavor symmetry group $G_F$. 
In this work we will consider only $G_F = SU(2)$, so our bounds will apply to any theory with a flavor symmetry that admits an $SU(2)$ subgroup.
We leave the study of other flavor symmetry groups for future work.

As anticipated, a powerful organizing principle to study the space of $\Nm = 2$ SCFTs are the unitarity bounds obtained from the underlying chiral algebra. It will be important to keep in mind that the central charges of any interacting $\Nm = 2$ SCFT satisfy~\cite{Beem:2013sza, Lemos:2015orc}:\footnote{The dual Coxeter number $h^\vee$ is $N$ for $SU(N)$, $N-2$ for $SO(N)$ and so on.}
\begin{align}
\label{eq:ck-bounds}
 k \ge \frac{24 c h^\vee}{12 c+ \text{dim} G_F}, \quad
 k(-180c^2 + 66 c + 3 \text{dim} G_F)+ 60c^2 h^\vee - 22c h^\vee \le 0.
\end{align}
For theories without flavor symmetry, the first inequality does not apply and the second one reduces to $c \ge 11/30$~\cite{Liendo:2015ofa}.

Perhaps the most familiar examples of $\Nm = 2$ SCFTs are Lagrangian models, the simplest one being the $SU(N)$ SYM theory coupled to an adjoint hypermultiplet.
This is precisely $\Nm = 4$ SYM theory, where part of the full $SU(4)_R$ symmetry is reinterpreted as an $SU(2)$ flavor symmetry.
Another interesting model is $\Nm = 2$ SCQCD, namely an $SU(N)$ SYM theory coupled to $N_f = 2N$ hypermultiplets.
The flavor symmetry is $SO(8)$ for $N=2$ and $SU(2N) \times U(1)$ for $N \ge 3$. Data associated to these theories is presented in table~\ref{tab:landscape-n2}.
When restricting to rank-one or equivalently $SU(2)$ gauge group, SCQCD saturates the two bounds in~\eqref{eq:ck-bounds}, while $\Nm = 4$ SYM saturates only the first one.

\begin{table}[h]
	\centering
	\renewcommand{\arraystretch}{1.5}
	\begin{tabular}{|c||c|c|c|c|c|}
		\hline
        & $H_0$ & $H_1$ & $H_2$ & $\Nm = 4$ SYM & $\Nm = 2$ SCQCD \\ \hline\hline 
		$G_F$               & 
		\;~ - \;~           & 
		$SU(2)$             & 
		$SU(3)$             & 
		$SU(2)$             & 
		$SU(2N)\times U(1)$ \\ \hline
		$\Delta_\varphi$    & 
		$\frac{6}{5}$       & 
		$\frac{4}{3}$       & 
		$\frac{3}{2}$       &
		$2, \ldots, N$      &
		$2, \ldots, N$      \\ \hline
		$c$                 & 
		$\frac{11}{30}$     & 
		$\frac{1}{2}$       & 
		$\frac{2}{3}$       &
		$\frac{N^2-1}{4}$   &
		$\frac{2N^2-1}{6}$  \\ \hline
		$k$                 & 
		-                   & 
		$\frac{8}{3}$       & 
		$3$                 &
		$N^2-1$             &
		$2N$                \\ \hline
	\end{tabular}
\caption{$\Nm=2$ theories that will appear in the discussion of our results.}
\label{tab:landscape-n2}
\end{table}

Three non-Lagrangian models that will play an important role in our analysis are the Argyres-Douglas theories listed in the first three columns of table \ref{tab:landscape-n2}. These models were originally discovered as fixed points of $\Nm=2$ $SU(2)$ supersymmetric QCD \cite{Argyres:1995xn}, and they correspond to vacua where a monopole and $N_f$ quarks become massless.\footnote{The $H_0$ theory had been found first in \cite{Argyres:1995jj} as a fixed point of $\Nm=2$ $SU(3)$ super Yang-Mills.} 
These theories have a rank-one Coulomb branch, and the flavor symmetry is $SU(N_f)$ where $N_f=1,2,3$ for $H_0$, $H_1$, $H_2$ respectively.
Moreover, they saturate the bounds~\eqref{eq:ck-bounds} and have therefore a distinguished position in the $\Nm=2$ landscape. 

These Argyres-Douglas models are isolated and strongly interacting with no standard Lagrangian description.\footnote{See however \cite{Maruyoshi:2016aim} for an interesting  approach based on susy enhancement along an $\Nm=1$ RG flow.} Despite this fact and thanks to superconformal symmetry, some aspects of these theories are under good analytic control. The central charges listed in the table were calculated using holography in \cite{Aharony:2007dj}, and their associated chiral algebras were conjectured in \cite{Buican:2015ina,Cordova:2015nma}. 
They can also be obtained as low energy theories on $D3$-branes probing $F$-theory singularities \cite{Aharony:1998xz}, a construction that naturally generalizes our models to higher ranks.
In addition to the flavor symmetry already discussed, the rank-$N$ models will enjoy an extra $SU(2)_L$ symmetry, which will have its own flavor central charge $k_L$.
The central charges of the higher rank models are~\cite{Aharony:2007dj}:
\begin{align}
 c &= \frac{1}{4} N^2 \Delta_\varphi + \frac{3}{4} N(\Delta_\varphi - 1) - \frac{1}{12}, \\
 k &= 2 N \Delta_\varphi, \\
 k_L &= N^2 \Delta_\varphi - N(\Delta_\varphi-1) - 1,
\end{align}
where $\Delta_\varphi$ is the dimension of the rank-one Coulomb branch generator in table~\ref{tab:landscape-n2}, and the remaining generators have dimensions $2\Delta_\varphi, \ldots, N\Delta_\varphi$.
This review of $\Nm = 2$ theories is by no means complete, but it will be sufficient for the discussion of our numerical results.

\subsection{Correlators, conformal blocks and selection rules}

Having reviewed the basic multiplets we are interested in, let us now have a brief review of superconformal kinematics. 
Conformal Ward identities imply that four-point functions depend on a function of two cross-ratios $z$ and $\zb$:
\begin{align}
 \langle \phi_i(x_1) \phi_j(x_2) \phi_k(x_3) \phi_l(x_4) \rangle
 = \frac{\Gm_{ijkl}(z,\bar z)}{x_{12}^{\Delta_i + \Delta_j} x_{34}^{\Delta_k + \Delta_l}}
   \left( \frac{x_{24}}{x_{14}} \right)^{\Delta_{ij}}
   \left( \frac{x_{14}}{x_{13}} \right)^{\Delta_{kl}}.
\end{align}
Using the Operator Product Expansion (OPE) in the $(12)\to(34)$ channel, one obtains the conformal block decomposition:
\begin{align}
 \Gm_{ijkl}(z,\bar z) 
 = \sum_{\Om \in \phi_i \times \phi_j} (-1)^\ell \lambda_{ij\bar \Om} \lambda_{kl\Om} \, 
        g_{\Delta,\ell}^{\Delta_{ij},\Delta_{kl}}(z,\bar z).
\end{align}
Here the sum runs only over conformal primaries that appear in the OPE $\phi_i \times \phi_j$, and the contribution of all conformal descendants is captured by the conformal blocks originally computed in~\cite{Dolan:2000ut,Dolan:2003hv}.
In our conventions, they are normalized as
\begin{align}
 g^{\Delta_{12},\Delta_{34}}_{\Delta,\ell}(z,\bar z) 
 = \frac{z \bar z}{z - \bar z} \Big(
     k^{\Delta_{12},\Delta_{34}}_{\Delta+\ell}(z)
     k^{\Delta_{12},\Delta_{34}}_{\Delta-\ell-2}(\bar z)
   - k^{\Delta_{12},\Delta_{34}}_{\Delta+\ell}(\bar z) 
     k^{\Delta_{12},\Delta_{34}}_{\Delta-\ell-2}(z)
 \Big),
\end{align}
where the one-dimensional blocks are of the familiar form:
\begin{align}
 k^{\Delta_{12},\Delta_{34}}_\beta(z) 
 = z^{\beta / 2} \,
   {}_2F_1 \Big( 
     \tfrac12(\beta-\Delta_{12}), \tfrac12(\beta+\Delta_{34}); \beta; z 
   \Big).
\end{align}
It will be understood from now on that $g_{\Delta,\ell} = g_{\Delta,\ell}^{0,0}$ and $k_\beta = k_\beta^{0,0}$.

When supersymmetry is present, the OPE coefficients of descendant operators can be related to the ones of the superprimary.
As a result, $\Gm_{ijkl}$ can be decomposed in terms of superconformal blocks
\begin{align}
 \Gm_{ijkl}(z,\bar z) 
 = \sum_{\Om \in \phi_i \times \phi_j} (-1)^\ell \lambda_{ij\bar \Om} \lambda_{kl\Om} \, 
        G_{\Delta,\ell}^{ij,kl}(z,\bar z),
\end{align}
where the sum now runs only over superprimary operators in the $\phi_i \times \phi_j$ OPE, and the superconformal blocks $G_{\Delta,\ell}^{ij,kl}$ are linear combinations of non-supersymmetric blocks.

In the rest of this section we will discuss the implications of $\Nm = 2$ supersymmetry on all possible correlation functions formed with (anti)chiral operators $\varphi$, $\bar \varphi$ and the moment map operator $M$.
Single correlators for chiral operators and moment maps were studied originally in  \cite{Beem:2014zpa}, while mixed correlators involving both types of operators have not been studied before. In order to bootstrap this system the first necessary step is to calculate the corresponding superconformal blocks, this is done in section~\ref{sec:blocks-chirals-moment-map} by imposing consistency between different $\Nm=1$ decompositions.

\subsubsection{Chiral correlators}
\label{sec:blocks-chirals}

We focus first on the four-point function of two chirals and two antichirals:
\begin{align}
 \langle \varphi_r(x_1) \varphi_r(x_2) \bar \varphi_{-r}(x_3) \bar \varphi_{-r}(x_4) \rangle
 = \frac{1}{x_{12}^{2\Delta_\varphi} x_{34}^{2\Delta_\varphi}}
   \sum_{\Om \in \varphi \times \varphi} |\lambda_{\varphi\varphi \bar \Om}|^2 \, 
        g_{\Delta,\ell}(z,\bar z).
\end{align}
To understand what operators appear in the sum we need to study the OPE $\varphi_r \times \varphi_r \sim \Om$. 
The non-supersymmetric selection rules require that only even spin operators with $2r$-charge appear in the sum.
Furthermore, the LHS is chiral and annihilated by the superconformal charge $S_\alpha$ (see~\cite{Poland:2010wg} for a proof), i.e. $\bar Q_{\dot \alpha} \Om = S_{\alpha} \Om = 0$.
Tables of supermultiplets can be found in~\cite{Cordova:2016emh}, where one starts with a primary operator at the top and all the $Q$ and $\Qb$-descendants are arranged in a diamond. The superselection rule implies that the only operator that contributes to the OPE is the one sitting in the right corner of this diamond.
Since only one operator in each multiplet contributes to the OPE, the superconformal blocks reduce to standard bosonic blocks and we define $G^{\varphi\varphi;\bar\varphi\bar\varphi}_{\Delta,\ell} \equiv g_{\Delta,\ell}$.

Going through the tables of superconformal multiplets, we obtain all operators that can appear in the OPE.
The resulting selection rule, together with the conformal blocks are summarized in table~\ref{tab:n2-phi-phi-OPE}. 

{
\renewcommand{\arraystretch}{1.5}
\renewcommand\tabcolsep{8pt}
\begin{table}[h]
\centering
\begin{tabular}{ | l | l | l | l | }
  \hline
  {\bf Multiplet} & 
  {\bf Block} & 
  {\bf Restrictions} \\ \hline \hline
  $L \bar B[0; 0]^{(0;2r)}$ & 
  $g_{2 \Delta_\varphi,0}$ & \\ \hline
  $L \bar A[\ell; \ell\text-2]^{(0;2r\text-2)}$ & 
  $g_{2 \Delta_\varphi + \ell,\ell}$ &
  $\ell \ge 2, \; \ell \; \text{even}$ \\ \hline \hline
  $L \bar B[0; 0]^{(2;2r\text-2)}$ & 
  $g_{2 \Delta_\varphi + 2,0}$ & \\ \hline
  $L \bar A[\ell; \ell\text-1]^{(1;2r\text-3)}$ & 
  $g_{2 \Delta_\varphi + \ell + 2,\ell}$ &
  $\ell \ge 2, \; \ell \; \text{even}$ \\ \hline
  $L \bar L[\ell; \ell]^{(0;2r\text-4)}_{\Delta-2}$ & 
  $g_{\Delta,\ell}$ & 
  $\Delta > 2 \Delta_\varphi + \ell + 2, \; 
   \ell \ge 0, \; 
   \ell \; \text{even}$ \\ \hline
\end{tabular} 
\caption{List of multiplets that appear in the $\varphi_r \times \varphi_r$ OPE, where $\varphi_r$ is the primary of $\Nm = 2$ chiral multiplet. For each multiplet, only one conformal descendant appears in the OPE, so we obtain non-supersymmetric bosonic blocks.}
\label{tab:n2-phi-phi-OPE}
\end{table}
}

Similarly, we can study the same four-point function with the operators in a different order:
\begin{align}
 \langle \varphi_r(x_1) \bar \varphi_{-r}(x_2) \varphi_r(x_3) \bar \varphi_{-r}(x_4) \rangle
 = \frac{1}{x_{12}^{2\Delta_\varphi} x_{34}^{2\Delta_\varphi}}
   \sum_{\Om \in \varphi \times \bar \varphi} |\lambda_{\varphi \bar\varphi \Om}|^2 \, 
        G_{\Delta,\ell}^{\varphi\bar\varphi;\varphi\bar\varphi}(z,\bar z).
\end{align}
Now the sum runs over all superprimaries in the OPE $\varphi_r \times \bar \varphi_{-r} \sim \Om$.
Using superconformal Ward identities it is easy to prove that only multiplets with vanishing $R$ and $r$ charge can appear. By going through the list of $\Nm = 2$ multiplets one obtains
\begin{align}
 \varphi_{r} \times \bar \varphi_{-r} 
 & \sim
   \mathds{1} 
 + A \bar A[\ell;\ell]^{(0;0)} 
 + L \bar L[\ell;\ell]_{\Delta}^{(0;0)}.
\end{align}
For each multiplet, all operators with $R = r = 0$ appear in $\varphi_r \times \bar \varphi_{-r}$, thus the superconformal blocks are linear combinations of bosonic blocks.
The superconformal block for the exchange of the long multiplet was originally computed in~\cite{Fitzpatrick:2014oza}, and takes a very compact form:
\begin{align}
\begin{split}
\label{eq:Gppbppb}
 G^{\varphi\bar\varphi;\varphi\bar\varphi}_{\Delta,\ell}(z,\bar z)
 & = (z \bar z)^{-1} g^{2,2}_{\Delta+2,\ell}(z,\bar z).
\end{split}
\end{align}
At the unitarity bound $\Delta = \ell + 2$, the $L\bar L$ multiplet shortens and we obtain the superconformal block associated to $A\bar A$.
In particular, the stress tensor belongs to $A\bar A[0;0]^{(0;0)}$, and all the $A\bar A[\ell;\ell]^{(0;0)}$ with $\ell \ge 1$ contain higher-spin conserved currents that are absent in interacting SCFTs.
Similarly, for $\ell=0$ and $\Delta=0$ we obtain the identity operator. 
Although not manifestly so, the above superconformal block can be expanded as a sum of bosonic blocks with $\Delta_{12} = 0$.
We do not need the full result, but let us note for future reference the contribution of the stress-tensor multiplet:
\begin{align} 
\begin{split}
\label{eq:ppbppb-stress-tensor}
 G^{\varphi\bar\varphi;\varphi\bar\varphi}_{2,0}(z, \zb)
 & =  g_{2,0}
   + \frac{1}{4}  g_{3,1}(z, \zb)
   + \frac{1}{60} g_{4,2}(z, \zb).
\end{split}
\end{align}
In order to study crossing symmetry, we will also need the following ordering:
\begin{align}
 \langle \varphi_r(x_1) \bar \varphi_{-r}(x_2) \bar \varphi_{-r}(x_3) \varphi_r(x_4) \rangle
 = \frac{1}{x_{12}^{2\Delta_\varphi} x_{34}^{2\Delta_\varphi}}
   \sum_{\Om \in \varphi \times \bar \varphi} |\lambda_{\varphi \bar\varphi \Om}|^2 \, 
        \tilde G_{\Delta,\ell}^{\varphi\bar\varphi;\varphi\bar\varphi}(z,\bar z).
\end{align}
The block $\tilde G$ is given by the same linear combination of non-supersymmetric blocks as $G$.
However, for each term in the sum, we must include a factor $(-1)^\ell$ depending on the spin of the exchanged operator.
All in all, the superconformal blocks in compact form are:
\begin{align}
\begin{split}
 \tilde G^{\varphi\bar\varphi;\bar\varphi\varphi}_{\Delta,\ell}(z,\bar z)
 & = (-1)^\ell (z \bar z)^{-1} g^{2,-2}_{\Delta+2,\ell}(z,\bar z).
\end{split}
\end{align}

The results of this section are summarized in table~\ref{tab:n2-phi-bphi-OPE}.

{
\renewcommand{\arraystretch}{1.5}
\renewcommand\tabcolsep{8pt}
\begin{table}[h]
\centering
\begin{tabular}{ | l | l | l | l | }
  \hline
  {\bf Multiplet} & 
  {\bf Block} & 
  {\bf Restrictions} \\ \hline \hline
  $\mathds 1$ & 
  $1$ &  \\ \hline
  $A\bar A[\ell; \ell]^{(0;0)}$ & 
  $G_{\ell + 2,\ell}^{\varphi \bar \varphi; \varphi \bar \varphi}$ & 
  $\ell \ge 0$ \\ \hline 
  $L \bar L[\ell; \ell]^{(0;0)}_{\Delta}$ & 
  $G_{\Delta,\ell}^{\varphi \bar \varphi; \varphi \bar \varphi}$ & 
  $\Delta > \ell + 2, \; \ell \ge 0$ \\ \hline
\end{tabular} 
\caption{List of multiplets that appear in the $\varphi_r \times \bar \varphi_{-r}$ OPE, where $\varphi_r$ is the primary of an $\Nm = 2$ chiral multiplet and $\bar \varphi_{-r}$ is its complex conjugate. The explicit form of the superconformal block is given in~\eqref{eq:Gppbppb}. The multiplets $A\bar A[\ell; \ell]^{(0;0)}$ for $\ell \ge 1$ contain higher-spin conserved currents and should be absent in an interacting SCFT.}
\label{tab:n2-phi-bphi-OPE}
\end{table}
}

\subsubsection{Moment map correlator}
\label{sec:blocks-moment-map}

Now we consider the four-point function of moment map operators.
In this work, we will restrict our attention to $G_F = SU(2)$, which could represent the full flavor symmetry of the theory or an $SU(2)$ subgroup. It is convenient to contract the $SU(2)_R$ indices with auxiliary vectors $t^I$ to unclutter the equations $M^A(x,t) = M^A_{IJ}(x) t^I t^J$. Using this notation the four-point function of moment maps can be decomposed into $SU(2)_R$ and flavor irreducible representations:
\begin{align}
\begin{split}
 \langle
   M^A(x_1, t_1)
   M^B(x_2, t_2)
&  M^C(x_3, t_3)
   M^D(x_4, t_4)
 \rangle \\
&= \frac{(t_1 \cdot t_2)^2 (t_3 \cdot t_4)^2}{x_{12}^4 x_{34}^4} 
   \sum_{R=0,2,4} 
   \sum_{i} P_i^{ABCD} P_R(y) \, a_{i,R}(z,\zb).
\end{split}
\end{align}
We contract the auxiliary vectors as $t_a \cdot t_b = \varepsilon_{IJ} t_a^I t_b^J$, and the $SU(2)_R$-invariant cross ratio is
\begin{align}
 w = \frac{(t_1 \cdot t_2)(t_3 \cdot t_4)}{(t_1 \cdot t_3)(t_2 \cdot t_4)}\,, \qquad
 y = \frac{2}{w} - 1\,.
\end{align}
Since the moment map is a triplet under $R$-symmetry, the four-point function decomposes into $[2] \otimes [2] = [0] \oplus [2] \oplus [4]$, where $[R]$ is the $(R+1)$-dimensional representation of $SU(2)$.
The projectors $P_R(y)$ are given by Legendre polynomials:
\begin{align}
 P_0(y) = 1\,, \qquad
 P_2(y) = y\,, \qquad
 P_4(y) = \frac12(3y^2 - 1)\,.
\end{align}
On the flavor symmetry side, we have an index $i$ which runs over all irreducible representations in the product of two adjoints $i \in \mathrm{ad}\, G_F \times \mathrm{ad}\, G_F$.
We use orthogonal projectors normalized as follows:
\begin{align}
 P_i^{ABCD} P_j^{DCEF} = \delta_{ij} P_i^{ABEF}\,, \qquad
 P_i^{ABBA} = \dim R_i\,.
\end{align}
For the case of interest to us $G_F = SU(2)$, the projectors are:
\begin{subequations}
 \begin{align}
 P_{\bf 1}^{ABCD} 
 & = \frac 13 \delta^{AB} \delta^{CD}\,, \\
 P_{\bf 3}^{ABCD} 
 & = \frac 12 (\delta^{AD} \delta^{BC} - \delta^{AC} \delta^{BD})\,, \\
 P_{\bf 5}^{ABCD} 
 & = \frac 12 (\delta^{AD} \delta^{BC} + \delta^{AC} \delta^{BD}) - P_{\bf 1}^{ABCD}\,.
\end{align}
\end{subequations}
A useful property of the moment map four-point function is that superconformal Ward identities relate the different $R$-symmetry channels.
In particular, the three $a_{i,R}$ for $R=0,2,4$ depend on a two-variable function $\Gm_i(z,\bar z)$ and a meromorphic function $f_i(z)$~\cite{Dolan:2001tt,Dolan:2004mu,Nirschl:2004pa}:
\begin{align}
\label{eq:aiR_mommap}
 a_{i,0}(z,\zb) & =
  \frac{2 z \zb - 3(z+\zb) + 6}{6} \Gm_i(z,\zb) -
  \frac{z\zb}{2(z-\bar z)}
  \left( \frac{(2-z) f_i(z)}{z} - \frac{(2-\bar z) f_i(\bar z)}{\bar z} \right), 
  \nonumber \\[0.5em]
 a_{i,2}(z,\zb) & =
  \frac{z \zb}{2(z - \bar z)}\big(f_i(z) - f_i(\bar z) \big) +
  \frac{z \zb - z - \zb}{2} \Gm_i(z,\zb), \\[0.5em]
 a_{i,4}(z,\zb) & = 
  \frac{z\zb}{6} \Gm_i(z,\zb)
  \nonumber .
\end{align}
Since the superconformal blocks satisfy the same Ward identities as the correlator, we 
can also express them in terms of $\Gm(z,\zb)$ and $f(z)$.
The selection rules for the moment map operator were first obtained in~\cite{Arutyunov:2001qw}, and the corresponding blocks were calculated in~\cite{Dolan:2001tt}.
We do not review the calculation here, but we just quote the result in table~\ref{tab:n2-M-M-OPE}.

{
\renewcommand{\arraystretch}{1.5}
\renewcommand\tabcolsep{8pt}
\begin{table}[h]
\centering
\begin{tabular}{ | l | l | l | l | }
  \hline
  {\bf Multiplet} & 
  {\bf Block $\Gm(u,v)$} & 
  {\bf Block $f(z)$} & 
  {\bf Restrictions} \\ \hline \hline
  $\mathds 1$ & 
  0 & 
  1 &
  \multicolumn{1}{c|}{$-$} \\ \hline
  $B \bar B[0; 0]^{(2;0)}$ & 
  $0$ & 
  $2 k_2$ &
  \multicolumn{1}{c|}{$-$} \\ \hline
  $B \bar B[0; 0]^{(4;0)}$ & 
  $6 u^{-1} g_{4,0}$ & 
  $6 k_4$ &
  \multicolumn{1}{c|}{$-$} \\ \hline
  $A \bar A[\ell; \ell]^{(0;0)}$ & 
  $0$ & 
  $-k_{2(\ell+2)}$ &
  $\ell \ge 0$ \\ \hline
  $A \bar A[\ell; \ell]^{(2;0)}$ & 
  $-2 u^{-1} g_{\ell+5,\ell+1}$ & 
  $-2 k_{2(\ell+3)}$ &
  $\ell \ge 0$ \\ \hline
  $L \bar L[\ell; \ell]^{(0;0)}_{\Delta}$ & 
  $u^{-1} g_{\Delta+2,\ell}$ & 
  $0$ &
  $\Delta > \ell + 2$ \\ \hline
\end{tabular} 
\caption{List of multiplets that appear in the $M \times M$ OPE, where $M$ is the $\Nm = 2$ moment map operator. The superconformal blocks can be expressed in terms of two functions $\Gm(z,\zb)$ and $f(z)$, as discussed around equation~\eqref{eq:aiR_mommap}. In our conventions, the contribution from the lowest-dimension operator is always unit normalized.}
\label{tab:n2-M-M-OPE}
\end{table}
}

\subsubsection{Chiral and moment map}
\label{sec:blocks-chirals-moment-map}

Finally, we consider the channel involving both chiral and moment map operators.
In this case, the superconformal blocks are not available in the literature.
Fortunately, we can leverage the knowledge of $\Nm = 1$ superblocks to easily obtain the required blocks.
The strategy is to build the $\Nm = 2$ superblocks as a linear combination of $\Nm = 1$ blocks, and by a mix of basic consistency conditions and $\Nm = 2$ selection rules, it turns out all free coefficients can be fixed.
We present the steps in detail in appendix~\ref{sec:app_blocks}.

First we consider the four-point function
\begin{align}
\label{eq:corr-PPbMM}
 \langle
        \varphi (x_1)
   \bar \varphi (x_2)
   M^A(x_3, t_3)
   M^B(x_4, t_4)
 \rangle
 = \frac{\delta^{AB} (t_3 \cdot t_4)^2}{|x_{12}|^{2\Delta_\varphi} |x_{34}|^4} 
   \sum_{\Om} \lambda_{\varphi\bar\varphi\Om} \lambda_{MM\Om}G^{\varphi\bar\varphi;MM}_{\Delta,\ell}(z,\bar z),
\end{align}
where the sum runs over all even-spin superprimaries which are both in the $\varphi_r \times\bar\varphi_{-r}$ and $M\times M$ OPEs (see tables~\ref{tab:n2-phi-bphi-OPE} and~\ref{tab:n2-M-M-OPE}).
Interestingly, the superconformal block can be written very compactly (see appendix~\ref{sec:app_blocks}):
\begin{align}
\label{eq:n2-block-ppbMM}
 G^{\varphi\bar\varphi;MM}_{\Delta,\ell}(z,\zb)
 = (z \bar z)^{-1} 
   g_{\Delta + 2,  \ell}^{2,0}(z,\zb).
\end{align}
As before, at the unitarity bound $\Delta = \ell + 2$ this block gives the contribution of the $A\bar A[\ell; \ell]^{(0;0)}$ multiplet.
In the appendix we give an expression for $G^{\varphi\bar\varphi;MM}_{\Delta,\ell}$ as a linear combination of non-supersymmetric blocks~\eqref{eq:block-ppbMM-ap}.
We do not need such an expression in general, only when $\Delta=2$, $\ell = 0$ to capture the contribution of the stress tensor:
\begin{align}
\label{eq:stress-tens-PPMM}
 G^{\varphi\bar\varphi;MM}_{2,0}(z, \zb)
 = g_{2,0} - \frac{1}{30} g_{4,2}(z,\zb).
\end{align}
These results are summarized in table~\ref{tab:n2-phi-bphi-mix-OPE}.

{
\renewcommand{\arraystretch}{1.5}
\renewcommand\tabcolsep{8pt}
\begin{table}[h!]
\centering
\begin{tabular}{ | l | l | l | l | }
  \hline
  {\bf Multiplet} & 
  {\bf Block} & 
  {\bf Restrictions} \\ \hline \hline
  $\mathds 1$ & 
  $1$ & \\ \hline
  $A\bar A[\ell; \ell]^{(0;0)}$ & 
  $G_{\ell + 2,\ell}^{\varphi \bar \varphi; MM}$ & 
  $\ell \ge 0, \; \ell \; \text{even}$ \\ \hline 
  $L \bar L[\ell; \ell]^{(0;0)}_{\Delta}$ & 
  $G_{\Delta,\ell}^{\varphi \bar \varphi; MM}$ & 
  $\Delta > \ell + 2, \; \ell \ge 0, \; \ell \; \text{even}$ \\ \hline
\end{tabular} 
\caption{List of multiplets that appear both in the $\varphi_r \times \bar \varphi_{-r}$ and $M \times M$ OPEs. For each multiplet, the superconformal block can be found in equation~\eqref{eq:n2-block-ppbMM}.}
\label{tab:n2-phi-bphi-mix-OPE}
\end{table}
}

Let us now consider the four-point function in the crossed channel
\begin{align}
\begin{split}
\label{eq:fourpt-PMMPb}
 \langle
        \varphi (x_1)
 &      M^A(x_2, t_2)
        M^B(x_3, t_3)
   \bar \varphi (x_4)
 \rangle \\
 & = \frac{\delta^{AB} (t_2 \cdot t_3)^2}
        {x_{12}^{\Delta_\varphi + 2} x_{34}^{\Delta_\varphi + 2}} 
   \left( \frac{x_{24}}{x_{14}} \right)^{\Delta_\varphi - 2}
   \left( \frac{x_{14}}{x_{13}} \right)^{\Delta_\varphi - 2}
   \sum_\Om |\lambda_{\varphi M \Om}|^2
            G^{\varphi M;M\bar\varphi}_{\Delta,\ell}(z,\bar z),
\end{split}
\end{align}
where the superconformal blocks derived in appendix~\ref{sec:app_blocks} are
\begin{align}
\begin{split}
\label{eq:block-PMMPb}
 G^{\varphi M;M\bar\varphi}_{\Delta,\ell}(z, \zb)
 & = (z \bar z)^{-1/2} 
   g_{\Delta + 2,  \ell}^{\Delta_\varphi - 1, 3 - \Delta_\varphi}(z, \zb).
\end{split}
\end{align}
The sum in~\eqref{eq:fourpt-PMMPb} runs over the superprimaries of three different multiplets.
For generic $\Delta$ it is a long multiplet $L\bar L$ and the block is given by~\eqref{eq:block-PMMPb}. 
At the unitarity bound $\Delta = \Delta_\varphi + \ell + 1$ we either obtain an $L\bar A$ multiplet if $\ell \ge 1$, or an $L\bar B$ multiplet for $\ell = 0$.

When we study crossing we will also need the blocks for the $\langle \varphi M \bar \varphi M \rangle$ ordering.
As before, we define them with a tilde:
\begin{align}
\begin{split}
 \tilde G^{\varphi M;M\bar\varphi}_{\Delta,\ell}(z, \zb)
 & = (-1)^\ell (z \bar z)^{-1/2}
   g_{\Delta + 2,  \ell}^{\Delta_\varphi - 1,  \Delta_\varphi - 3}(z, \zb).
\end{split}
\end{align}
The results in this section are summarized in table~\ref{tab:n2-phi-M-OPE}.

{
\renewcommand{\arraystretch}{1.5}
\renewcommand\tabcolsep{8pt}
\begin{table}[h!]
\centering
\begin{tabular}{ | l | l | l | l | }
  \hline
  {\bf Multiplet} & 
  {\bf Block} & 
  {\bf Restrictions} \\ \hline \hline
  $L\bar B[0;0]^{(2;r)}$ & 
  $G_{\Delta_\varphi+1,0}^{\varphi M; M\bar \varphi}$ & \\ \hline
  $L\bar A[\ell;\ell\text-1]^{(1;r\text-1)}$ & 
  $G_{\Delta_\varphi+\ell+1,\ell}^{\varphi M; M\bar \varphi}$ &
  $\ell \ge 1$ \\ \hline
  $L \bar L[\ell; \ell]^{(0;r\text-2)}_{\Delta}$ & 
  $G_{\Delta,\ell}^{\varphi M; M\bar \varphi}$ & 
  $\Delta > \Delta_\varphi + \ell + 1, \; \ell \ge 0$ \\ \hline
\end{tabular} 
\caption{List of multiplets that appear in the $\varphi_r \times M$ OPE. For each multiplet, the superconformal block can be found in equation~\eqref{eq:block-PMMPb}.}
\label{tab:n2-phi-M-OPE}
\end{table}
}


\section{Crossing equations}
\label{sec:crossing}

With the selection rules and superconformal blocks at hand, we are finally ready to present the crossing equations of interest.
Although it is generally an easy exercise to obtain them, for the moment map four-point function one needs to take into account the contributions coming from the chiral algebra. We will review the most important results which are derived in more detail in~\cite{Beem:2014zpa}.

\subsection{Generalities}

The bootstrap for four-point functions of different scalars was first studied in~\cite{Kos:2014bka}.
As usual, one demands that the OPE decomposition in the $(12)\to(34)$ channel is equivalent to the $(14)\to(23)$ channel:\footnote{There will also be flavor symmetry indices which for simplicity we do not consider yet.}
\begin{align}
 \contraction{\langle}{\phi_i}{(x_1)}{\phi_j}
 \contraction{\langle \phi_i(x_1) \phi_j(x_2)}{\phi_k}{(x_3)}{\phi_l}
 \langle \phi_i(x_1) \phi_j(x_2) \phi_k(x_3) \phi_l(x_4) \rangle
 = 
 \contraction[2ex]{\langle}{\phi_i}{(x_1) \phi_j(x_2) \phi_k(x_3)}{\phi_l}
 \contraction[1ex]{\langle \phi_i(x_1)}{\phi_j}{(x_2)}{\phi_k}
 \langle \phi_i(x_1) \phi_j(x_2) \phi_k(x_3) \phi_l(x_4) \rangle\,.
\end{align}
Upon expanding in conformal blocks, this implies two independent crossing equations
\begin{align}
\label{eq:general-cross-eq}
 \sum (-1)^\ell \lambda_{ij\Om} \lambda_{kl\Om} E_{\pm,\Delta,\ell}^{ij,kl}(z,\bar z)
 \pm
 \sum (-1)^\ell \lambda_{kj\Om} \lambda_{il\Om} E_{\pm,\Delta,\ell}^{kj,il}(z,\bar z)
 = 0\,,
\end{align}
where we have defined
\begin{align}
\begin{split}
\label{eq:Epm}
 E_{\pm,\Delta,\ell}^{ij,kl}(z,\bar z)
 = & (z-\bar z) \bigg[ 
     (z \bar z)^{-\frac{\Delta_i + \Delta_j}{2}} 
     g_{\Delta,\ell}^{\Delta_{ij},\Delta_{kl}}(z,\bar z) \\
 & \mp \big( (1-z) (1-\bar z) \big)^{-\frac{\Delta_i + \Delta_j}{2}} 
     g_{\Delta,\ell}^{\Delta_{ij},\Delta_{kl}}(1-z,1-\bar z) 
 \bigg]\,.
\end{split}
\end{align}
We have multiplied by $(z-\bar z)$ to simplify the approximation of blocks in terms of polynomials, as explained in appendix~\ref{sec:app_numerics}.
In what follows, we will use the same notation for superblocks, for example $E^{MM;\varphi\bar\varphi}$ is obtained from~\eqref{eq:Epm} with $\Delta_{1,2} = 2$, $\Delta_{3,4} = \Delta_\varphi$ and $g^{\Delta_{12},\Delta_{34}}_{\Delta,\ell} \to G^{MM;\varphi\bar\varphi}_{\Delta,\ell}$.
\footnote{We proceed similarly for the superblocks with tilde. For example $\tilde E^{\varphi M;M\bar\varphi}$ is obtained from~\eqref{eq:Epm} with $\Delta_{2,3} = 2$, $\Delta_{1,4} = \Delta_\varphi$ and $g^{\Delta_{12},\Delta_{34}}_{\Delta,\ell} \to \tilde G^{\varphi M;M\bar\varphi}_{\Delta,\ell}$.}

\subsection{Chiral correlators}
\label{sec:crossing-chirals}

The constraints imposed by crossing symmetry on a system of $\Nm = 1$ chiral correlators in $4d$ was first studied in~\cite{Poland:2010wg} and later improved in~\cite{Poland:2011ey}. 
The analogous system with $\Nm = 2$ supersymmetry has been studied by~\cite{Beem:2014zpa,Lemos:2015awa,Cornagliotto:2017snu}.
Applying~\eqref{eq:general-cross-eq} to $\langle \varphi \bar \varphi \varphi \bar \varphi\rangle$ and $\langle \varphi \varphi \bar \varphi \bar \varphi \rangle$ one obtains three independent equations
\begin{align}
\label{eq:cross-chirals}
 \vec I_{c} 
 + \sum_{\Om \in A^+}
   | \lambda_{\varphi \bar \varphi \Om} |^2 \, \vec U_{\Delta,\ell}
 + \sum_{\Om \in A^-}
   | \lambda_{\varphi \bar \varphi \Om} |^2 \, \vec V_{\Delta,\ell}
 + \sum_{\Om \in B^+}
   | \lambda_{\varphi \varphi \bar \Om} |^2 \, \vec W_{\Delta,\ell}
  = 0,
\end{align}
where
\begin{align}
 \vec U_{\Delta,\ell} 
 =
 \begin{pmatrix}[1.3]
          E_{+,\Delta,\ell}^{\varphi \bar \varphi; \varphi \bar \varphi} \\
   \tilde E_{+,\Delta,\ell}^{\varphi \bar \varphi; \varphi \bar \varphi} \\
   \tilde E_{-,\Delta,\ell}^{\varphi \bar \varphi; \varphi \bar \varphi}
 \end{pmatrix},
 \qquad
 \vec V_{\Delta,\ell} 
 =
 \begin{pmatrix}[1.3]
          E_{+,\Delta,\ell}^{\varphi \bar \varphi; \varphi \bar \varphi} \\
   \tilde E_{+,\Delta,\ell}^{\varphi \bar \varphi; \varphi \bar \varphi} \\
   \tilde E_{-,\Delta,\ell}^{\varphi \bar \varphi; \varphi \bar \varphi}
 \end{pmatrix}, \qquad
 \vec W_{\Delta,\ell} 
 =
 \begin{pmatrix}[1.3]
   0 \\
     E_{+,\Delta,\ell}^{\varphi \varphi; \bar \varphi \bar \varphi} \\
   - E_{-,\Delta,\ell}^{\varphi \varphi; \bar \varphi \bar \varphi}\,.
 \end{pmatrix}.
\end{align}
Furthermore, we have separated the contribution of the unit operator and the stress tensor into
\begin{equation}
 \vec I_c = \vec U_{0,0} + \frac{\Delta_{\varphi}^2}{6c} \vec U_{2, 0}\,.
\end{equation}
The normalization of the stress tensor is easily obtained from~\eqref{eq:ppbppb-stress-tensor} and the requirement that it is normalized as $\sim \frac{\Delta_\varphi^2}{360c} g_{4,2}$ (for details see \cite{Poland:2010wg,Poland:2011ey}).
The ranges of the sums in~\eqref{eq:cross-chirals} can be read from tables~\ref{tab:n2-phi-phi-OPE} and~\ref{tab:n2-phi-bphi-OPE}:
\begin{align}
\label{eq:rangesAAB}
\begin{split}
 A^+ & = \left\{ 
    \ell \ge 0, \;
    \ell \text{ even}, \;
    \Delta \ge \ell + 2
 \right\}, \\
 A^- & = \left\{ 
    \ell \ge 0, \;
    \ell \text{ odd}, \;
    \Delta \ge \ell + 2
 \right\}, \\
 B^+ & = \left\{ 
    \ell \ge 0, \;
    \ell \text{ even}, \;
    \Delta = 2\Delta_\varphi + \ell
 \right\} \cup \left\{ 
    \ell \ge 0, \;
    \ell \text{ even}, \;
    \Delta \ge 2\Delta_\varphi + \ell + 2
 \right\}.
\end{split}
\end{align}
The operators in the $\varphi \times \bar \varphi$ OPE are divided into even and odd spins $A^\pm$. 
Even though the distinction is not necessary at this point, we keep it in analogy to the mixed system in section~\ref{sec:crossing-mixed}.

\subsection{Moment map correlator}
\label{sec:crossing-moment-map}

The crossing equations for the moment map are a bit more intricate than for a regular non-supersymmetric four-point function.
Let us remember that this correlator is completely determined in terms of two variable functions $\Gm_i(z,\bar z)$ and meromorphic functions $f_i(z)$.
The index $i \in \mathrm{ad}\, G_F \times \mathrm{ad}\, G_F$ runs over representations that appear in the product of two flavor adjoint representations.
The projectors $P_i^{ABCD}$ transform under crossing as:
\begin{align}
 P_i^{ABCD} = F_i^{\ph ij} P_j^{CBAD}, \qquad
 F_{SU(2)} = 
 \begin{pmatrix}[1.3]
  \frac13 & \ph- \frac13 & \ph- \frac13 \\
  1       & \ph- \frac12 &    - \frac12 \\
  \frac53 &    - \frac56 & \ph- \frac16
 \end{pmatrix}.
\end{align}
With this definition it is a simple exercise to show that the crossing relations for $f_i(z)$ decouple from the rest, and take the simple form
\begin{align}
\label{eq:chiral-algebra-cross}
 f_i \left( \frac{z}{z-1} \right)
 = (-1)^{\text{symm}(i)} f_i(z), \qquad
 F_j^{\ph ji} f_j(z)
 = \left( \frac{z}{z-1} \right)^2 f_i(1-z)\,.
\end{align}
Here $\text{symm}(i)$ is $0$ for representations that are symmetric under the exchange of points $1\leftrightarrow 2$, and $1$ for the antisymmetric ones.
For the case of $SU(2)$, $\text{symm}({\bf 1}) = \text{symm}({\bf 5}) = 0$ and $\text{symm}({\bf 3}) = 1$.
Interestingly, the crossing equations~\eqref{eq:chiral-algebra-cross} determine $f_i(z)$ up to a free parameter.
Alternatively, one can obtain $f^{ABCD}(z)$ as the four-point function of affine Kac-Moody currents of the chiral algebra associated to the $\Nm = 2$ theory~\cite{Beem:2013sza}.
Using either procedure, one obtains
\begin{subequations}
\label{eq:fi-SU2}
\begin{align}
 f_{\bf 1}(z) 
 & = \frac{3 - 6z + (5-\frac8k) z^2 - (2-\frac8k) z^3 + z^4}{(1-z)^2}, \\
 f_{\bf 3}(z) 
 & = \frac{-\frac8k z +\frac{12}{k} z^2 + (2-\frac4k) z^3 - z^4}{(1-z)^2}, \\
 f_{\bf 5}(z) 
 & = \frac{(2+\frac4k) (z^2-z^3) + z^4}{(1-z)^2}.
\end{align}
\end{subequations}
The free parameter $k$ is the flavor central charge that appears in the four-point function of flavor currents.

Now one can expand the $f_i(z)$ in terms of one-dimensional blocks, and extract the corresponding OPE coefficients. 
Using table~\ref{tab:n2-M-M-OPE} these can be mapped to the OPE coefficients in the $\Gm(z,\bar z)$ block expansion.\footnote{In order to do this mapping unambiguously one needs to assume the absence of higher-spin currents, or equivalently, that $A\bar A[\ell;\ell]^{(0;0)}$ for $\ell \ge 1$ are absent in our equations.}
With this information, one can resum the contribution of the $B \bar B[0;0]^{(4;0)}$ and $A \bar A[\ell;\ell]^{(2;0)}$ multiplets to $\Gm(z,\bar z)$, which splits into protected and unprotected pieces:
\begin{align}
   \Gm_i(z, \bar z) 
 = \Gm_i^{\text{sh}}(z, \zb)
 + \Gm_i^{\text{long}}(z, \zb), \qquad
 \Gm_i^{\text{long}}(z, \zb)
 = \sum |\lambda_{MM\Om_i}|^2 \, u^{-1} g_{\Delta,\ell}(z,\zb).
\end{align}
The conformal block decomposition of the long (unprotected) piece involves only $L\bar L[\ell,\ell]^{(0;0)}_{\Delta}$ multiplets, and the corresponding block appears in table~\ref{tab:n2-M-M-OPE}.
For $SU(2)$ flavor, the short (protected) pieces were computed in~\cite{Beem:2014zpa}:
\begin{align}
\label{eq:Gshort-SU2}
 \Gm^{\text{sh}}_{\mathbf 1}(z,\bar z) 
&= \frac{\log \left(1-\bar{z}\right)}{z-\bar{z}}
   \left(
     \frac{6}{c}+
     \frac{8 z^2}{k (1-z)}
     -\frac{z^2 \left(z^2-2 z+2\right)}{(1-z)^2}
   \right) \nonumber \\
& \quad - \frac{\log (1-z)}{z-\bar{z}}
   \left(
     \frac{6}{c}+
     \frac{8 \bar{z}^2}{k \left(1-\bar{z}\right)}-
     \frac{\bar{z}^2 \left(\bar{z}^2-2 \bar{z}+2\right)}{\left(1-\bar{z}\right)^2}
   \right) \nonumber \\
& \quad - \frac{6 \log (1-z) \log \left(1-\bar{z}\right)}{c z \bar{z}}, \\[0.5em]
 \Gm^{\text{sh}}_{\mathbf 3}(z,\bar z) 
 & = \frac{(2-z) z \log \left(1-\bar{z}\right)}{\left(z-\bar{z}\right) (1-z)}
   \left(\frac{4}{k}-\frac{z^2}{1-z}\right)
 - \frac{\left(2-\bar{z}\right) \bar{z} \log (1-z)}{\left(z-\bar{z}\right)   
         \left(1-\bar{z}\right)}  
   \left(\frac{4}{k}-\frac{\bar{z}^2}{1-\bar{z}}\right), \nonumber \\[0.5em]
 \Gm^{\text{sh}}_{\mathbf 5}(z,\bar z) 
 & = 
 - \frac{z^2 \log \left(1-\bar{z}\right)}{\left(z-\bar{z}\right) (1-z)}
   \left(\frac{4}{k}+\frac{z^2-2 z+2}{1-z}\right)
 + \frac{\bar{z}^2 \log (1-z)}{\left(z-\bar{z}\right) \left(1-\bar{z}\right)}
   \left(\frac{4}{k}+\frac{\bar{z}^2-2 \bar{z}+2}{1-\bar{z}}\right)\,. \nonumber
\end{align}

Keeping this information in mind, we are finally ready to write a set of crossing equations that constrain the unprotected spectrum of $\Nm = 2$ theories.
Besides~\eqref{eq:chiral-algebra-cross}, crossing symmetry implies
\begin{align}
\label{eq:crossing-H}
 \left( F_j^{\ph ji} \pm \delta_j^{\ph ji} \right) 
 \big( \Hm_j(z,\bar z) \pm \Hm_j(1-z,1-\bar z) \big)
 \equiv
 \left( F_j^{\ph ji} \pm \delta_j^{\ph ji} \right) 
 \Hm_{\pm,j}(z,\bar z)
 = 0\,,
\end{align}
where
\begin{align}
\label{eq:H-def}
 \Hm_i(z,\bar z) 
 = (z - \bar z) (z \bar z)^{-2} \Gm_i(z,\bar z)
 - \frac{1}{2(z \bar z)^2} \left( 
   \frac{z}{z-1} f_i(\bar z) - \frac{\bar z}{\bar z-1} f_i(z)
 \right)\,.
\end{align}
Out of the six crossing equations~\eqref{eq:crossing-H}, only three of them are independent.
When we expand them in blocks, there will be a piece which corresponds to the long operators, for which $f_i(z)$ drops out and $\Gm_i(z,\zb) \to u^{-1} g_{\Delta+2,\ell}$, see table~\ref{tab:n2-M-M-OPE}.
On the other hand, for the protected part both~\eqref{eq:fi-SU2} and~\eqref{eq:Gshort-SU2} will be relevant, and we collect them in the vector $\vec I_{c,k}$.
All in all, we get
\begin{align}
\label{eq:cross-MMMM}
\begin{split}
 \vec I_{c,k}
 + \sum_{A^+}         |\lambda_{MM\Om_{\bf 1}}|^2 \, \vec U_{\Delta, \ell}
 + \sum_{C_{\bf 3}^-} |\lambda_{MM\Om_{\bf 3}}|^2 \, \vec X_{\Delta, \ell}
 + \sum_{C_{\bf 5}^+} |\lambda_{MM\Om_{\bf 5}}|^2 \, \vec Y_{\Delta, \ell}
 = 0\,.
\end{split}
\end{align}
with
\begin{align}
\begin{split}
 \quad
 \vec U_{\Delta, \ell}
 =
 \begin{pmatrix}[1.3]
  \ph- 4 \, E^{MM,MM}_{+,\Delta,\ell} \\
  \ph- 2 \, E^{MM,MM}_{+,\Delta,\ell} \\
     - 2 \, E^{MM,MM}_{-,\Delta,\ell}
 \end{pmatrix},
 \;\;
 \vec X_{\Delta, \ell}
 =
 \begin{pmatrix}[1.3]
  3 \, E^{MM,MM}_{+,\Delta,\ell} \\
  9 \, E^{MM,MM}_{+,\Delta,\ell} \\
  3 \, E^{MM,MM}_{-,\Delta,\ell}
 \end{pmatrix},
 \;\;
 \vec Y_{\Delta, \ell}
 =
 \begin{pmatrix}[1.3]
  \ph- 5 \, E^{MM,MM}_{+,\Delta,\ell} \\
     - 5 \, E^{MM,MM}_{+,\Delta,\ell} \\
  \ph- 5 \, E^{MM,MM}_{-\Delta,\ell}
 \end{pmatrix}.
\end{split}
\end{align}
For convenience we have defined 
\begin{align}
 E^{MM,MM}_{\pm,\Delta,\ell}(z,\zb)
 = (z-\zb) \Big[ 
   (z\zb)^{-3} g_{\Delta,\ell}(z, \zb)
   \mp \big( (1-z)(1-\zb) \big)^{-3} g_{\Delta,\ell}(1-z, 1-\zb)
 \Big].
\end{align}
The contribution of all the protected operators depends only on $c$ and $k$.
It can be constructed by combining the definition of $\Hm$ in~\eqref{eq:H-def} with~\eqref{eq:fi-SU2} and~\eqref{eq:Gshort-SU2}:
\begin{align}
\begin{split}
 \vec I_{c,k} =
 \begin{pmatrix}[1.3]
  \ph- 4 \Hm_{+,\mathbf 1}^\text{sh} 
     + 3 \Hm_{+,\mathbf 3}^\text{sh} 
     + 5 \Hm_{+,\mathbf 5}^\text{sh} \\
  \ph- 2 \Hm_{+,\mathbf 1}^\text{sh}
     + 9 \Hm_{+,\mathbf 3}^\text{sh}
     - 5 \Hm_{+,\mathbf 5}^\text{sh} \\
     - 2 \Hm_{-,\mathbf 1}^\text{sh}
     + 3 \Hm_{-,\mathbf 3}^\text{sh}
     + 5 \Hm_{-,\mathbf 5}^\text{sh}
 \end{pmatrix}.
\end{split}
\end{align}
Regarding the ranges of summation in~\eqref{eq:cross-MMMM}, we have three different channels depending on the flavor symmetry representation.
The singlet channel is equivalent to $A^+$ defined in~\eqref{eq:rangesAAB}, while the ${\bf 3}$ and ${\bf 5}$ channels give:
\begin{align}
\begin{split}
 C_{\bf 3}^- & = \left\{ 
    \ell \ge 0, \;
    \ell \text{ odd}, \;
    \Delta > \ell + 2
 \right\}, \\
 C_{\bf 5}^+ & = \left\{ 
    \ell \ge 0, \;
    \ell \text{ even}, \;
    \Delta > \ell + 2
 \right\}.
\end{split}
\end{align}

\subsection{Full mixed system}
\label{sec:crossing-mixed}

The crossing equations for the mixed system are a simple extension of the ones presented so far.
By considering $\langle \varphi M \bar \varphi M \rangle$ and $\langle \varphi \bar \varphi M M \rangle$ we obtain three extra  equations and a new channel $\varphi \times M$.
The full system is then
\begin{align}
\label{eq:crossing-full}
\begin{split}
 \vec I_{c,k}
 + \sum_{\Om \in A^+}
   \left( \lambda_{M\!M\Om}^* \; \lambda_{\varphi\bar\varphi \Om}^* \right)
   \vec U_{\Delta,\ell}
   \begin{pmatrix} \lambda_{M\!M\Om} \\ \lambda_{\varphi\bar\varphi \Om} \end{pmatrix}
&+ \sum_{\Om \in A^-}
   | \lambda_{\varphi\bar\varphi \Om} |^2
   \vec V_{\Delta,\ell}
 + \sum_{\Om \in B^+}
   | \lambda_{\varphi\varphi \Om} |^2
   \vec W_{\Delta,\ell} \\
 + \sum_{\Om \in C^-_{\bf 3}}
   | \lambda_{M\!M \Om} |^2
   \vec X_{\Delta,\ell}
&+ \sum_{\Om \in C^+_{\bf 5}}
   | \lambda_{M\!M \Om} |^2
   \vec Y_{\Delta,\ell}
 + \sum_{\Om \in D^\pm}
   | \lambda_{\varphi M \Om} |^2
   \vec Z_{\Delta,\ell}
 = 0.
\end{split}
\end{align}
The explicit crossing vectors are easy to obtain and are given in appendix~\ref{sec:cross-vectors}.
In the new channel the summation range can be obtained from table~\ref{tab:n2-phi-M-OPE}:
\begin{align}
\begin{split}
 D^\pm & = \left\{ 
    \ell \ge 0, \;
    \Delta \ge \Delta_\varphi + \ell + 1 
 \right\}.
\end{split}
\end{align}
We will often refer to the different channels by the name of the crossing vector.
For example, the multiplets that appear in the $\varphi\times M$ OPE will be referred to as $Z$-channel multiplets, and similarly for the other channels.


\section{Numerical bounds}
\label{sec:numerics}

In this section we use the numerical bootstrap of~\cite{Rattazzi:2008pe} to obtain bounds on the space of $\Nm = 2$ superconformal theories. 
Arguably the most important numerical bootstrap result is the precise determination of the critical exponents of the three-dimensional Ising and $O(N)$ models~\cite{ElShowk:2012ht,El-Showk:2014dwa,Kos:2014bka,Simmons-Duffin:2015qma,Kos:2015mba,Kos:2016ysd,Chester:2019ifh}.
In the supersymmetric literature, one can find studies of $3d$ models with minimal supersymmetry~\cite{Atanasov:2018kqw,Rong:2018okz,Rong:2019qer},  $\Nm=2$ supersymmetry~\cite{Bobev:2015jxa,Chester:2015lej,Chester:2015qca,Bae:2016jpi,Baggio:2017mas} and
maximal $\Nm = 8$ supersymmetry~\cite{Chester:2014fya,Chester:2014mea,Agmon:2017xes,Agmon:2019imm}.
Similarly, in four dimensions there have been studies of $\Nm=1$ theories~\cite{Poland:2010wg,Poland:2011ey,Berkooz:2014yda,Poland:2015mta,Li:2017ddj},
$\Nm = 2$ theories~\cite{Beem:2014zpa,Lemos:2015awa,Cornagliotto:2017snu},
$\Nm = 3$ theories~\cite{Lemos:2016xke} and of
$\Nm = 4$ SYM theory~\cite{Beem:2013qxa,Beem:2016wfs}.
Finally there have also been bootstrap studies of supersymetric theories in two~\cite{Lin:2015wcg,Lin:2016gcl,Cornagliotto:2017dup}, five~\cite{Chang:2017cdx} and six~\cite{Beem:2015aoa,Chang:2017xmr} dimensions and for supersymmetric defects~\cite{Liendo:2018ukf,Gimenez-Grau:2019hez}.
A mixed correlator between Coulomb and moment map operators similar in spirit to our work has been studied in $3d$ $\Nm = 4$ theories in \cite{Chang:2019dzt}.
A pedagogical introduction to the modern numerical bootstrap is~\cite{Chester:2019wfx}.

\subsection{Chiral correlators}

In this section we focus exclusively on chiral correlators. Some of our results are new, while others are improved versions of the ones previously obtained in~\cite{Beem:2014zpa,Lemos:2015awa,Cornagliotto:2017snu}. For details of the numerical implementation for this and all subsequent sections see appendix~\ref{sec:app_numerics}.

\subsubsection{OPE bounds and spectrum}

\begin{figure}[htb]
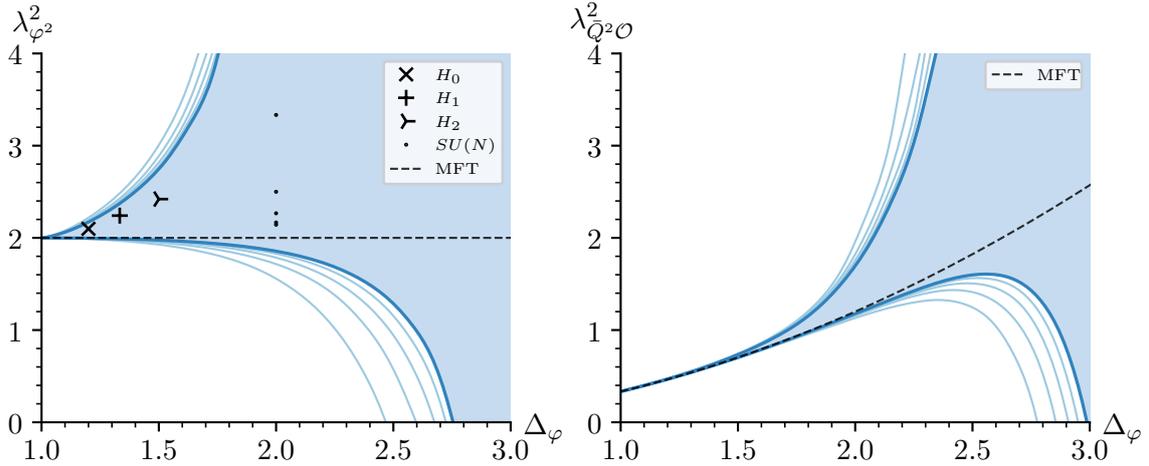

\centering
\begin{tabular}{@{}c@{}c@{}}
 \input{./figures/n2Chirals_opeBoundP2_deltaP.pgf} &
 \input{./figures/n2Chirals_opeBoundLA2_deltaP.pgf}
\end{tabular}
\caption{Upper and lower bounds on the OPE coefficients of operators $\varphi^2$ and $\Qb^2 \Om$ in $\varphi \times \varphi$, where $\varphi$ is the highest weight of a chiral multiplet. Curves are shown for $\Lambda = 16, 20, 24, 28, 32$, and the allowed region is filled. In the left figure, the values for the rank one $H_0$, $H_1$ and $H_2$ theories were computed in the large $r$-charge limit in~\cite{Hellerman:2018xpi,Grassi:2019txd}.
We also present free $SU(N)$ gauge theory values for $N = 2, \ldots, 6$, which as $N \to \infty$ approach the mean-field theory solution (MFT).}
\label{fig:n2Chirals_opeBoundP2_deltaP}
\end{figure}

As summarized in table~\ref{tab:n2-phi-phi-OPE}, the OPE of two chiral fields $\varphi_r \times \varphi_r$ contains a family of protected operators captured by conformal blocks $g_{2\Delta_\varphi+\ell,\ell}$ for $\ell = 0, 2, 4, \ldots$.
These operators should be interpreted as double traces of the chiral primary operators $\varphi \, \partial^{\mu_1} \ldots \partial^{\mu_\ell} \varphi$.
For $\ell = 0$ this gives another Coulomb branch operator that we denote $\varphi^2$, while for $\ell = 2$ the exchanged operator is a level-two descendant in its multiplet which we denote $\Qb^2 \Om$.
In figure~\ref{fig:n2Chirals_opeBoundP2_deltaP} we plot upper and lower bounds on the OPE coefficients $\lambda_{\varphi^2}$ and $\lambda_{\Qb^2 \Om}^2$ as a function of the dimension of the external field $\Delta_\varphi = r/2$, without making any assumptions about the spectrum.
Notice that upper and lower bounds are close to each other for low values of the external dimension $\Delta_{\varphi}$. Luckily, the Argyres-Douglas models listed in table \ref{tab:landscape-n2} are characterized by a low value of $\Delta_{\varphi}$, which means the bounds are particularly useful to constrain these theories.
The equivalent bounds for $\ell \ge 4$ are qualitatively equal to the $\ell = 2$ case, in the sense that for low values of $\Delta_\varphi$ the OPE coefficient must be approximately given by their MFT values.

It is interesting now to compare our results with the works~\cite{Hellerman:2018xpi,Grassi:2019txd}, where analytic expressions were obtained for $\lambda_{\varphi^2}^2$ for the Argyres-Douglas theories of table \ref{tab:landscape-n2}. 
In \cite{Hellerman:2018xpi} the authors obtained an expression valid to all orders in $1/r$, without taking into account exponentially-suppressed non-perturbative contributions, which could be relevant for the case of interest to us where $r = \Delta_\varphi$ is of order one.\footnote{We thank D. Orlando for an interesting discussion on this point.}
The approach of \cite{Grassi:2019txd} also relies on taking a large $r$-charge limit, that can be analyzed using Random Matrix Theory (RMT) techniques. 
The two methods give very similar results, which we have added to figure~\ref{fig:n2Chirals_opeBoundP2_deltaP}, and observe that they sit in the allowed region. Figure~\ref{fig:n2Chirals_opeBoundP2_deltaP} is agnostic regarding the value of the central charge $c$, so we can refine these bounds by fixing $c$ to the Argyres-Douglas values of table \ref{tab:landscape-n2}, and also by increasing the number of derivatives $\Lambda$. This allows for a better comparison between the bootstrap and large $r$-charge results which we present in table~\ref{tab:OPE-bounds}. There is surprising agreement between the two approaches, especially considering the analytic results were calculated using a large-$r$ expansion, and the values of $r$ for these operators are quite low. The values for the $\Qb^2 \Om$ OPE coefficients are exclusive to the bootstrap and have not been estimated by other techniques.

{
\renewcommand{\arraystretch}{1.5}
\renewcommand\tabcolsep{8pt}
\begin{table}[htb]
\centering
\begin{tabular}{ | c | c | c | c | c | }
  \hline
                    & 
  {\bf Lower bound} & 
  {\bf Upper bound} &
  {\bf Resummed~\cite{Hellerman:2018xpi} } &
  {\bf RMT~\cite{Grassi:2019txd} } \\ \hline \hline
$\lambda^2_{\varphi^2}$ for $H_0$ &
$2.142596$ &
$2.16509$ &
$2.1181$ &
$2.0982$  \\ \hline
$\lambda^2_{\varphi^2}$ for $H_1$ &
$2.216735$ &
$2.35462$ &
$2.2129$ &
$2.2412$  \\ \hline
$\lambda^2_{\varphi^2}$ for $H_2$ &
$2.299679$ &
$2.69898$ &
$2.3457$ &
$2.4206$  \\ \hline
\hline
$\lambda^2_{\bar Q^2 \Om}$ for $H_0$ &
$0.468394$ &
$0.46893$ &
$-$ &
$-$  \\ \hline
$\lambda^2_{\bar Q^2 \Om}$ for $H_1$ &
$0.571321$ &
$0.57544$ &
$-$ &
$-$  \\ \hline
$\lambda^2_{\bar Q^2 \Om}$ for $H_2$ &
$0.714878$ &
$0.73218$ &
$-$ &
$-$  \\ \hline
\end{tabular} 
\caption{Upper and lower bounds on the OPE coefficients of operators $\varphi^2$ and $\bar Q^2 \Om$ in $\varphi \times \varphi$, where $\varphi$ is the highest weight of a chiral multiplet. The parameters $\Delta_\varphi$ and $c$ are fixed to the known values of the rank-one Argyres-Douglas theories in table~\ref{tab:landscape-n2}.
All bounds have been obtained at $\Lambda = 50$.
In the rightmost columns we compare with the values computed by resuming an expansion in $1/r$ to all orders~\cite{Hellerman:2018xpi}, and using Random Matrix Theory in~\cite{Grassi:2019txd}.}
\label{tab:OPE-bounds}
\end{table}
}

There is a second, less obvious motivation for looking at these bounds. Using the extremal functional method it is possible to extract the spectrum of the theory which lives at the boundary of the exclusion region~\cite{ElShowk:2012hu}.
This idea has been used successfully in many applications, most importantly the $3d$ Ising model~\cite{El-Showk:2014dwa,Simmons-Duffin:2016wlq}.
In our case we have two numerical bounds that are quite close to each other, and the hope is that the extracted spectrum is a good approximation of the actual spectrum of the Argyres-Douglas models. The results of the analysis are collected in table~\ref{tab:spectrum} in the appendix.
We are focusing on the $\ell = 0$ operators in the $U$ and $W$ channels.
In particular, $\Delta_U$ is the dimension of the first superprimary in the $\varphi \times \bar \varphi$ OPE, $\Delta_U'$ of the second superprimary, and so on.
For the $\varphi \times \varphi$ OPE, $\Delta_W$ and $\Delta_W'$ denote the dimension of the exchanged operators rather than the superprimaries.
For the rank-one $H_0$ and $H_1$ theories a summary of results is
\begin{equation}
\label{eq:approx-spec-H01}
\begin{alignedat}{6}
 &  H_0 \; : \quad
 && \Delta_U   \sim 2.7, \quad
 && \Delta_U'  \sim 5.9, \quad
 && \Delta_U'' \sim 9.2, \quad
 && \Delta_W   \sim 4.8, \quad
 && \Delta_W'  \sim 7,   \\[0.5em]
 &  H_1 \; : \quad
 && \Delta_U   \sim 3.0,     \quad
 && \Delta_U'  \sim 5.9,     \quad
 && \Delta_U'' \sim 9-13,    \quad
 && \Delta_W   \sim 5.3,     \quad
 && \Delta_W'  \sim 6-8.
\end{alignedat}
\end{equation}
These are rough averages of the results in table~\ref{tab:spectrum}, for which there is no rigorous way to estimate the errors.
In the next section we are going to use these numbers as a guide and attempt to isolate these models using bounds on scaling dimensions, and as we will see, a consistent picture emerges.

\subsubsection{Dimension bounds}

\begin{figure}[htb]
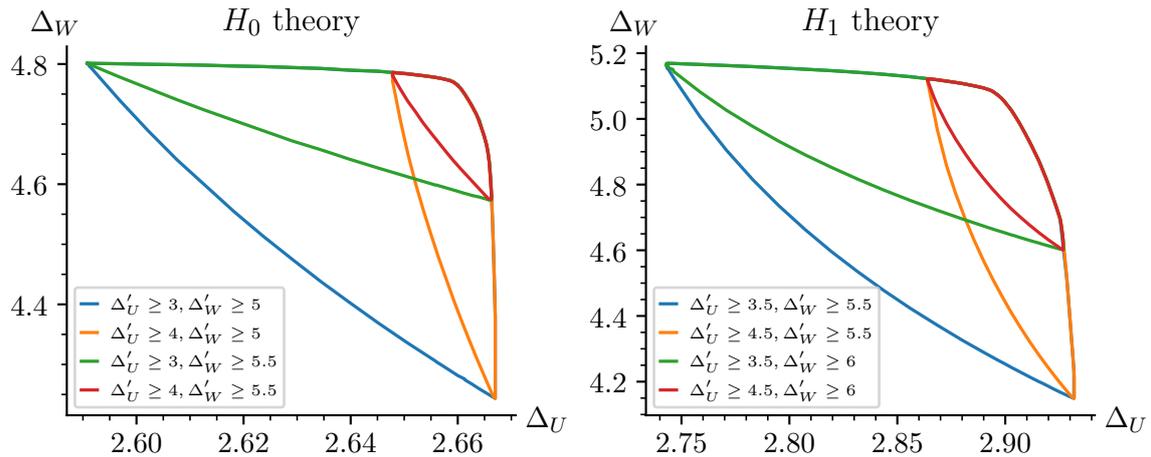

\centering
\begin{tabular}{@{}c@{}c@{}}
 \input{./figures/gapVW_H0_cont.pgf} &
 \input{./figures/gapVW_H1_cont.pgf}
\end{tabular}
\caption{Dimension $\Delta_U$ of the lowest dimension unprotected operator in the $\varphi \times \bar \varphi$ channel, versus dimension $\Delta_W$ of the lowest dimension unprotected operator in the $\varphi \times \varphi$ channel. We fix the dimension of the external operator $\varphi$ and the central charge $c$, and impose gaps with the next operators in the spectrum $\Delta_U'$ and $\Delta_W'$. All bounds have been obtained with cutoff $\Lambda = 32$.}
\label{fig:gapVW_cont}
\end{figure}

Having obtained a rough estimate of the spectrum of operators for the $H_0$ and $H_1$ rank-one Argyres-Douglas theories, let us now try to isolate them using the numerical bootstrap for conformal dimensions. More precisely, we will fix the dimensions of the external chiral operator $\varphi$ and the central charge to $c$  to the values listed in table \ref{tab:landscape-n2}.
We then plot the allowed region for $\Delta_U$ and $\Delta_W$, assuming they are the dimensions of the lowest-lying operators in their respective channels. We also assume gaps for the next operator in the spectrum consistent with~\eqref{eq:approx-spec-H01}:
\begin{equation}
\label{eq:gapsUW-H01}
\begin{alignedat}{3}
 &  H_0 \; : \quad
 && \Delta_U' \ge \{ 3.0, 4.0 \}, \quad
 && \Delta_W' \ge \{ 5.0, 5.5 \},   \\
 &  H_1 \; : \quad
 && \Delta_U' \ge \{ 3.5, 4.5 \}, \quad
 && \Delta_W' \ge \{ 5.5, 6.0 \}.
\end{alignedat}
\end{equation}
The results of this analysis can be found in figure~\ref{fig:gapVW_cont}.
We observe that even for the most conservative choices of gaps, the allowed region is a fairly small island in the $(\Delta_U, \Delta_W)$ plane. The effect of reducing the gaps in either channel is to allow for solutions to crossing with smaller dimensions.

\begin{figure}[htb]
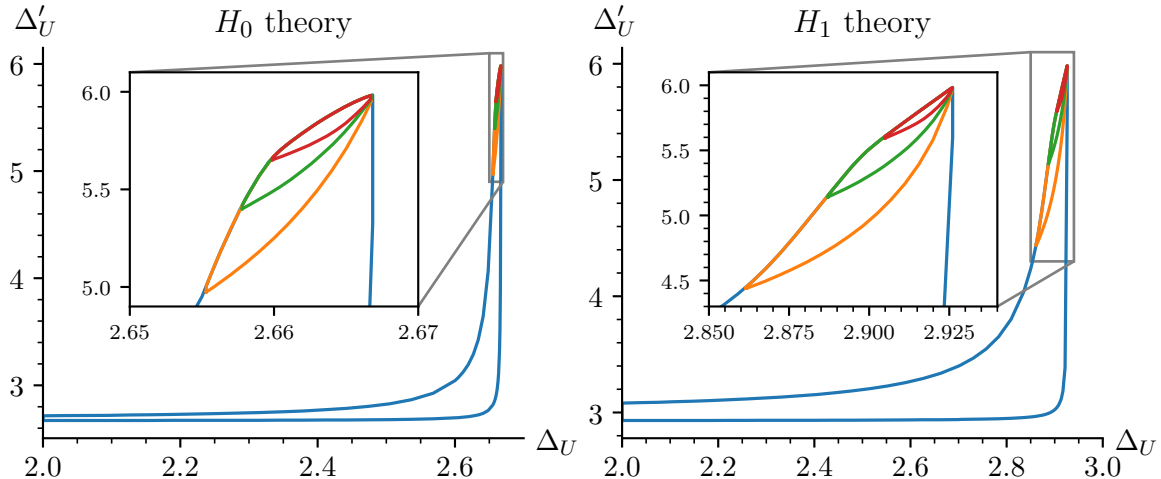

\centering
\begin{tabular}{@{}c@{}c@{}}
 \input{./figures/allowedDimPPb_deltaP_2gap_H0_cont.pgf} &
 \input{./figures/allowedDimPPb_deltaP_2gap_H1_cont.pgf}
\end{tabular}
\caption{Dimensions $\Delta_U$ versus $\Delta_U'$ of the two lowest dimension unprotected operators in the $\varphi \times \bar \varphi$ channel. We fix the dimension of the external operator $\varphi$ and the central charge $c$, and impose a gap with the next operator in the spectrum $\Delta_U'' \ge 6, 7, 8, 9$. For $\Delta_U''$ greater than 7 we provide a zoom to the allowed region, which form small islands isolated from the continuum. All bounds have been obtained with cutoff $\Lambda = 32$.}
\label{fig:allowedDimPPb_deltaP_2gap_cont}
\end{figure}

An alternative but similar strategy is to focus on a particular channel, like the $U$-channel that contains operators neutral under all symmetries.  We plot the allowed values of $\Delta_U$ and $\Delta_U'$, assuming they are the lowest dimension operators, and assume different gaps for the next operator $\Delta_U''$ consistent with the approximate spectrum in~\eqref{eq:approx-spec-H01}:
\begin{equation}
\label{eq:gaps2U-H01}
  H_0, \, H_1 \; : \quad
 \Delta_U'' \ge \{ 6.0, \, 7.0, \, 8.0, \, 9.0 \}.
\end{equation}
The results of this analysis can be found in figure~\ref{fig:allowedDimPPb_deltaP_2gap_cont}.
In this case we observe two qualitatively distinct behaviors.
For $\Delta_U'' \ge 7, 8, 9$ once again we obtain a small allowed region in the the $(\Delta_U,\Delta_U')$ plane.
The estimate for $\Delta_U$ from this analysis is compatible with the one from figure~\ref{fig:gapVW_cont}.
On the other hand, when the gap is lowered further to $\Delta_U'' \ge 6$, the allowed region is no longer a small disconnected island.
Let us focus on the $H_0$ plot, although the conclusion is identical for $H_1$.
There are clearly two different regimes when $\Delta_U'' \ge 6$.
In the first, $\Delta_U$ can take any value between the unitarity bound 2 and $\sim 2.6$, provided that $\Delta_U' \sim 2.66$.
In the second, $\Delta_U'$ can take any value between $\sim 2.7$ and $\sim 6$ provided that $\Delta_U \sim 2.66$.
Summarizing, we are assuming there are only two operators with $\Delta_U,\Delta_U' < \Delta_U''$.
If the gap $\Delta_U''$ is small enough, crossing allows one operator to have arbitrary conformal dimension as long as the other is at $\Delta_U \sim 2.66$.
Thus, the bootstrap insists on having a long operator with dimension $\Delta_U \sim 2.66$, even when the gaps are lowered, but it cannot resolve the position of the next operator $\Delta_U'$.

We should point out that these results are not on the same footing as the Ising model island \cite{Kos:2014bka}, where the gaps were physically justified by assuming only two relevant operators in the spectrum, and moreover the island was obtained by studying mixed correlators. Our analysis was inspired by the one in \cite{Li:2017kck}, where numerical islands were obtained using a single-correlator bootstrap, by assuming mild gaps around conserved currents, similar to what we did here. 

We expect this single-correlator approach to give at least qualitative results for the $H_0$ theory, which we are assuming is in a sense ``simple'' and perhaps one of the models that has the best chance to be solved by bootstrap methods. Circumstantial evidence in favor of its simplicity include the fact that the theory is rank one, it has no Higgs branch, it has the minimum allowed value of $c$ among interacting theories, and the associated chiral algebra is the Yang-Lee edge singularity, arguably the simplest non-trivial $2d$ model with Virasoro symmetry.

For $H_1$ on the other hand we expect more structure. This model does have some simplifying features similar to those of $H_0$,  however there is an $SU(2)$ flavor symmetry and therefore Higgs branch operators. 
We will therefore consider the full mixed system of correlators in our attempts to corner this theory. 

\subsection{Moment map correlators}
\label{sec:moment-map-bounds}

\begin{figure}[htb]
\centering
\input{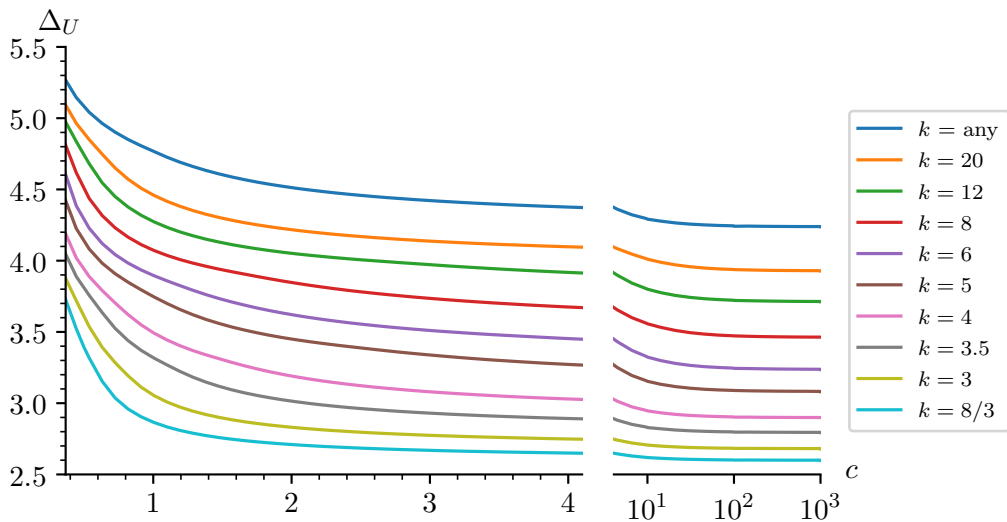}
\caption{Upper bound on the dimension $\Delta_U$ of the first singlet operator in the OPE $M \times M$ of two moment map operators. 
The values of the flavor central charge from bottom to top are $k = 8/3, 3, 7/2, 4, 5, 6, 8, 12, 20, \text{any}.$
The bounds for $c = 10^3$ and $c = \infty$ are identical with the resolution of the plot.
All results are obtained at $\Lambda = 32$. 
}
\label{fig:allowedDimMMSing_c}
\end{figure}

Before we jump to the full mixed system, let us also look at the single correlator of moment map operators, see~\eqref{eq:cross-MMMM}.
For later purposes, we will be mostly be interested on bounding the dimension of operators which are neutral under all symmetries, i.e. multiplets appearing in the $U$ channel.
These bounds depend heavily on the values of the central charge $c$ and the flavor central charge $k$.
In figure~\ref{fig:allowedDimMMSing_c} we present upper bounds on $\Delta_U$, the dimension of the first unprotected operator in the singlet channel of the $M \times M$ OPE.
We have chosen the values of $k$ to match some of the theories in table~\ref{tab:landscape-n2}, for example $H_1$, $H_2$ and $SU(N)$ SCQCD, and the rest provide a convenient interpolation between them.

We see that the upper bound on $\Delta_U$ decreases as we increase the central charge $c$, while it increases as we increase $k$.
If we do not make assumptions about $c$, the upper bound is dominated by the small $c$ values.
If we do not make assumptions about $k$, the bound is dominated by the large $k$ region.
These considerations will be important when we look at the mixed correlator results.
A more comprehensive survey of bounds from the moment map four-point function is available in~\cite{Beem:2014zpa}, including dimension bounds on the $X$ and $Y$ channel, as well as bounds on $c$ and $k$.

\subsection{Mixed correlators}

We are finally ready to consider the full mixed system of crossing equations~\eqref{eq:crossing-full}.
The bounds we derive no longer apply to the rank-one $H_0$ theory, because there is no flavor symmetry and therefore no moment map operator.
However, they apply to the higher rank version of $H_0$ with the flavor given by the $SU(2)_L$ symmetry, as well as all other theories discussed in section~\ref{ref:case-studies}.

In general one expects the mixed correlator bootstrap to be most efficient when there is an overlap between the two types of operators.
For this reason we focus in the $U$ channel, which appears both in the $\varphi \times \bar \varphi$ and $M \times M$ OPEs, and also in the $Z$ channel, which contains operators in the $\varphi \times M$ OPE. This last channel is of particular interest because it captures a new family of operators inaccessible from the single correlators studied so far.
Numerical bounds on the central charges $c$ and $k$ can also be obtained, however the analytic bounds obtained from the associated chiral algebra~\cite{Beem:2013sza,Lemos:2015orc} are quite strong, and the numerical methods used here will not improve on them.

\subsubsection{\texorpdfstring{$U$}{U}-channel dimension bounds}
\label{sec:U-mixed-dimension-bounds}

We start bounding the dimension $\Delta_U$ of the first unprotected multiplet in the $\varphi \times \bar \varphi$ and $M \times M$ OPEs.
The crossing equations~\eqref{eq:crossing-full} contain three canonical parameters:
the central charge $c$, the flavor central charge $k$ and the dimension $\Delta_\varphi$ of the Coulomb branch operator.
Ideally we would like to obtain $\Delta_U$ for all possible values of $(c,k,\Delta_\varphi)$, but exploring a three-dimensional parameter space is computationally intensive.
Instead, in figure~\ref{fig:allowedDimPPb_deltaP_noK} we plot $\Delta_U$ for different values of $(c, \Delta_\varphi)$ without restrictions on $k$, 
and in figure~\ref{fig:allowedDimPPb_deltaP_noC} we plot $\Delta_U$ for different values of $(k, \Delta_\varphi)$ without restrictions on $c$.

Before we discuss the results in detail, let us summarize some of our expectations. 
The single correlator bound of $\Delta_U$ as a function of $\Delta_\varphi$ using the chiral correlators was studied in detail in~\cite{Beem:2014zpa,Lemos:2015awa,Cornagliotto:2017snu}.
The upper bound on $\Delta_U$ grows with $\Delta_\varphi$ with approximately the same slope as the mean field theory solution $\Delta_U \sim 2 \Delta_\varphi$.
On the other hand, from the moment map correlator the bound is $2.5 \le \Delta_U \le 5.5$, which is independent of $\Delta_\varphi$ and changes with $c$ and $k$ as shown in figure~\ref{fig:allowedDimMMSing_c}.
It is natural to expect that for small $\Delta_\varphi$ the upper bound is dominated by the chiral bound, while for large $\Delta_\varphi$ it is dominated by the moment map bound.
This is indeed the behavior we observe in figures~\ref{fig:allowedDimPPb_deltaP_noK} and~\ref{fig:allowedDimPPb_deltaP_noC}.
We also know that for $c,k \to \infty$ the numerical bounds cannot rule out the intersection of the mean field theory (MFT) solutions:
\begin{align}
\label{eq:mft}
 \Delta_U = 2\Delta_\varphi \quad \text{for} \quad \Delta_\varphi \le 2, \qquad
 \Delta_U = 2\Delta_M = 4   \quad \text{for} \quad \Delta_\varphi \ge 2.
\end{align}
These are plotted with a black dashed line in the figures.

\begin{figure}[htb]
\centering
\input{./figures/allowedDimPPb_deltaP_noK.pgf}
\caption{Upper bound on the dimension $\Delta_U$ of the first unprotected operator in the $\varphi \times \bar \varphi$ and $M\times M$ OPEs as a function of the dimension of the chiral operator $\Delta_\varphi$.
The flavor central charge $k$ is not fixed to a particular value.
In the top blue curve, the central charge $c$ is also free, while it takes values $c = 1/2, 1, 2, \infty$ in the curves below it.
The black dashed curve is the mean field theory value~\eqref{eq:mft}.
All results are computed with $\Lambda = 24$.}
\label{fig:allowedDimPPb_deltaP_noK}
\end{figure}

In figure~\ref{fig:allowedDimPPb_deltaP_noK} the flavor central charge $k$ is arbitrary, and as discussed in section~\ref{sec:moment-map-bounds}, the upper bound will be dominated by the $k \to \infty$ region of parameter space.
Let us first look at the $\Delta_\varphi \le 2$ regime. For low values of the central charge, the upper bound insists on staying above the mean-field theory (MFT) solution even when $\Lambda$ is increased.
For large values of the central charge the upper bound gets closer to the MFT value.
As expected, when we approach $\Delta_\varphi = 1$ the upper bound approaches the free-theory value $\Delta_U = 2$ regardless of the central charge.
The behavior is significantly different for $\Delta_\varphi \ge 2$,  where the curves start flattening as dictated by the moment map crossing symmetry.
For small values of the central charge $1/2 \le c \le 1$ this transition is smooth. However, as we increase the central charge $c \ge 2$ the transition becomes sharp resulting in a kink around $\Delta_\varphi \sim 2$.

By extrapolating the numerics to $\Lambda \to \infty$ it seems that the kink of the $c = \infty$ curve will eventually land on $(\Delta_\varphi,\Delta_U) = (2, 4)$ which corresponds to mean field theory (MFT). The conclusion is then that for large central charge the numerical bootstrap rules out any theory with a leading gap larger than the MFT value. This is precisely what was observed in $\Nm=4$ SYM in \cite{Beem:2013qxa,Beem:2016wfs}. In this case, the mean field theory solution is interpreted as the large $N$ and large $\lambda=g_{YM}^2 N$ limit of $\Nm=4$ SYM, whose correlators are captured by tree-level supergravity. 
We now have a similar phenomenon but for $\Nm=2$ SCFTs with $SU(2)$ flavor symmetry, which incidentally also includes $\Nm=4$ SYM. 
If we consider the $\Nm=2$ decomposition of $\Nm=4$ theories, part of the $SU(4)$ $R$-symmetry gets re-interpreted as a global $SU(2)$ flavor symmetry.
Furthermore, the decomposition of the ${\bf 20'}$ multiplet (the one studied in \cite{Beem:2013qxa,Beem:2016wfs}) into $\Nm = 2$ contains a chiral and antichiral operator of dimension $\Delta_\varphi = 2$, and a moment map multiplet~\cite{Dolan:2002zh}:
\begin{align}
	\Om_{{\bf 20'}} \sim M + \varphi + \bar \varphi + \ldots,
\end{align}
the same multiplets that we are studying in this work.
Figure~\ref{fig:allowedDimPPb_deltaP_noK} presents a bound on the lowest-dimension multiplet which has a scalar superprimary that is neutral under $R$-symmetry, so it should include $\Nm=4$ SYM. The plot is however more general and valid for any $\Nm=2$ theory with $SU(2)$ flavor. 
For $2 \le c < \infty$ it is unclear to us whether the kinks corresponds to physical $\Nm = 2$ theories.

\begin{figure}[htb]
\centering
\input{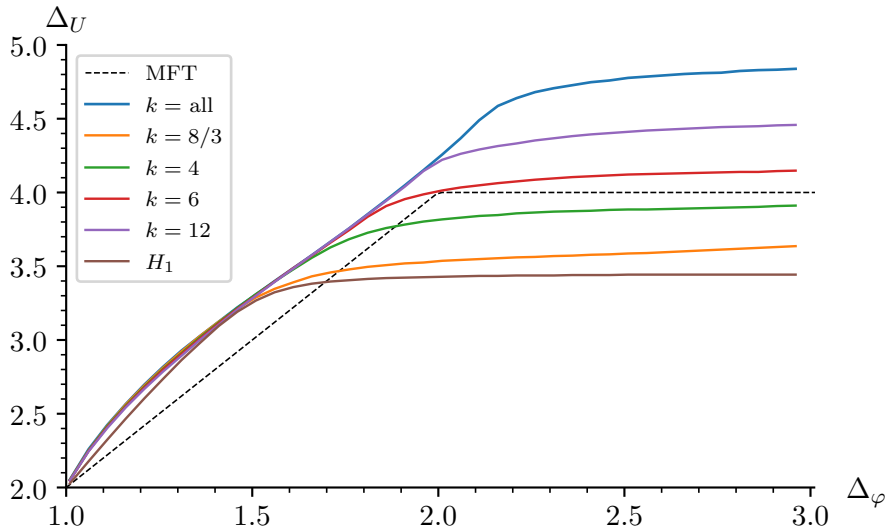}
\caption{Upper bound on the dimension $\Delta_U$ of the first unprotected operator in the $\varphi \times \bar \varphi$ and $M\times M$ OPEs as a function of the dimension of the chiral operator $\Delta_\varphi$.
The central charge $c$ is not fixed to a particular value.
In the top blue curve, the flavor central charge $k$ is also unfixed, while it takes values $k = 8/3, 4, 6, 12$ in the curves below it.
For the brown curve, both central charges are fixed to the values of the $H_1$ theory $c = \frac{1}{2}$, $k = \frac{8}{3}$.
The black dashed curve is the mean field theory value~\eqref{eq:mft}.
All results are computed with $\Lambda = 24$.}
\label{fig:allowedDimPPb_deltaP_noC}
\end{figure}

In figure~\ref{fig:allowedDimPPb_deltaP_noC} we consider the reversed situation, the central charge $c$ is not fixed, but we allow $k$ to take different values.
In this case, we observe a behavior qualitatively similar as before.
For $\Delta_\varphi \le 1.5$, the upper bound stays above MFT and grows almost parallel to it.
For $1.5 \le \Delta_\varphi \le 2.2$, depending on the value of $k$, the curves start to be dominated by the moment map part of crossing and they flatten.
This flattening is smooth and we do not observe any kinks like the ones found before.
The lowest upper bound in figure~\ref{fig:allowedDimPPb_deltaP_noC} is obtained by fixing both central charges $c$ and $k$ to the values of the $H_1$ theory.
For $H_1$ we know $\Delta_\varphi = \frac{4}{3}$, which corresponds to an upper bound $\Delta_U = 2.94$.
For larger $\Delta_\varphi$ the upper bound becomes flat at the value $\Delta_U = 3.4$ for $\Lambda = 24$.
As one increases the number of derivatives $\Lambda$, the upper bound $\Delta_U = 2.94$ is very robust and does not decrease, while the asymptotic value $\Delta_U = 3.4$ has not converged and is still decreasing.
Ideally, for $\Lambda \to \infty$ the two upper bounds would coincide, in which case the point $(\Delta_\varphi, \Delta_U) = (1.33, 2.94)$ would become a kink, and one could claim that the  $H_1$ theory saturates the numerical bounds.
Looking at the asymptotics of our numerics however it seems unlikely $\Delta_U = 3.4$ will go down below $\Delta_U \sim 3$.

\subsubsection{Focusing on \texorpdfstring{$H_1$}{H1}}
\label{sec:H1-mixed-dimension-bounds}

Let us now change gears for a moment. Instead of presenting bounds applicable to general $\Nm = 2$ SCFTs, we will try to use crossing symmetry to focus on the rank-one $H_1$ Argyres-Douglas theory.
Besides the central charges $c$, $k$ and the dimension of the Coulomb branch generator $\Delta_\varphi$, using the superconformal index~\cite{Buican:2015ina,Song:2015wta} one can show that the short multiplet $L\bar B[0;0]^{(2;r)}$ that could appear in the $\varphi \times M$ OPE is missing\footnote{We thank L. Rastelli for suggesting that this OPE coefficient might vanish in the $H_1$ theory, and J. Song for confirming that this is indeed the case.} (see table~\ref{tab:n2-phi-M-OPE}).
One should be aware that this is not a definitive proof that the multiplet is absent, and it could be a consequence of cancelations between different contributions to the index\footnote{We thank Madalena Lemos for comments regarding this issue.}.
We will nevertheless assume the multiplet is absent, and try to leverage this information to learn about the spectrum of the $H_1$ theory.

\begin{figure}[htb]
\centering
\input{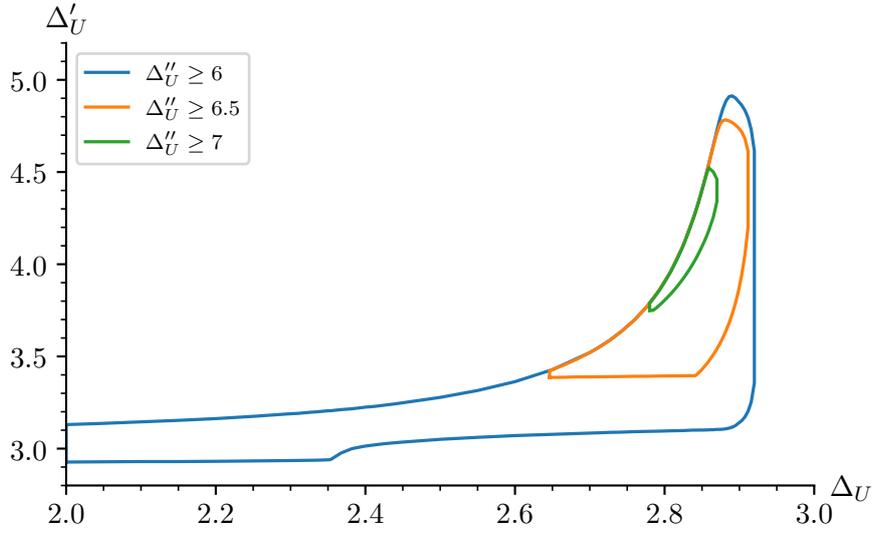}
\caption{Allowed values of $\Delta_U$ versus $\Delta_U'$, the dimensions of the first two multiplets in the $\varphi \times \bar \varphi$ and $M \times M$ OPEs. 
We assume that these are the only two operators below $\Delta_U'' \ge \{6, 6.5, 7\}$ and the parameters $\Delta_\varphi$, $c$, $k$ are fixed to the values of the rank-one $H_1$ theory.
The allowed region is enclosed between the upper and lower curves. 
The numerical optimization is using $\Lambda = 24$.}
\label{fig:mixed_2gap_H1_cont}
\end{figure}

Following the approach of figure~\ref{fig:allowedDimPPb_deltaP_2gap_cont}, we assume that there are only two operators in the $U$ channel with $\Delta_U,\Delta_U' \le \Delta_U''$, and find the allowed region in the plane $(\Delta_U, \Delta_U')$.
Using only the single correlator and assuming a sparse spectrum we managed to obtain a small island. Now we will complement that result by adding constraints coming from Higgs branch operators which we know are present in the $H_1$ theory.
Using the full mixed correlator system we observe in figure~\ref{fig:mixed_2gap_H1_cont} that the gaps $\Delta_U'' \ge \{6, 6.5, 7\}$ are allowed, but larger gaps are ruled out.
For the smallest gap we find a change of behavior in the lower bound around $\Delta_U = 2.35$, for which we do not currently have an interpretation.
To be on the safe side, in the next figure we will assume the most conservative gap $\Delta_U'' \ge 6$.

\begin{figure}[htb]
\centering
\input{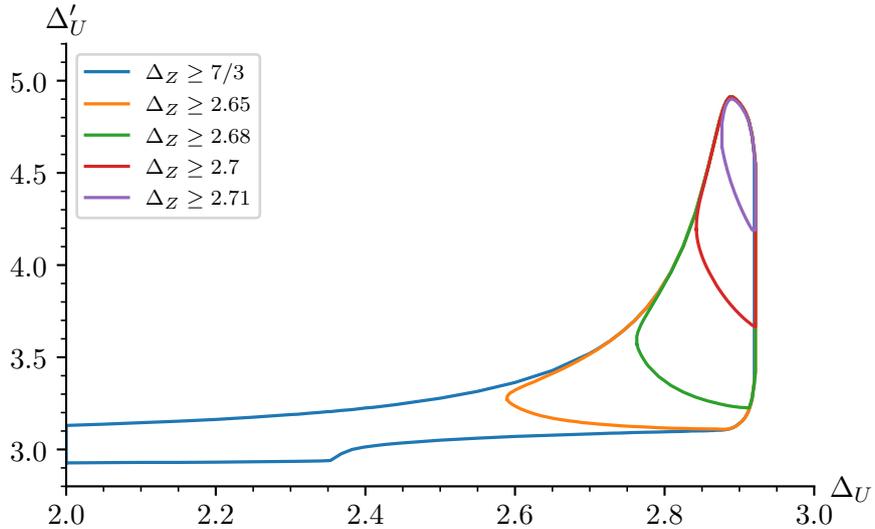}
\caption{Allowed values of $\Delta_U$ versus $\Delta_U'$, the dimensions of the first two multiplets in the $\varphi \times \bar \varphi$ and $M \times M$ OPEs, as a function of the gap $\Delta_Z$ in the channel $\varphi \times M$.
We assume that these are the only two operators below $\Delta_U'' \ge 6$, that the short multiplet $\varphi M$ is missing, and the parameters $\Delta_\varphi$, $c$ and $k$ are fixed to the values of the rank-one $H_1$ theory.
The allowed region is enclosed between the upper and lower curves. 
The numerical optimization is using $\Lambda = 24$.}
\label{fig:mixed_2gap_gapZ_H1_cont}
\end{figure}

The advantage of having the mixed correlator system is that we have a host of gap combinations we can assume.
In particular, since we know that the short operator $\varphi M \in \varphi \times M$ is missing in the $H_1$ theory, it is natural to impose a gap in this channel until the first long operator $\Delta_Z$.
The results, shown in figure~\ref{fig:mixed_2gap_gapZ_H1_cont}, indicate that as we increase the gap in the $Z$ channel, the allowed region shrinks to a small island around $(\Delta_U, \Delta_U') \sim (2.9, 5)$.
In this case the value of $\Delta_U$ is saturating the upper bound in figure~\ref{fig:allowedDimPPb_deltaP_noC}.
We are tempted to conjecture that $H_1$ is characterized by a solution of crossing without the short multiplet $\varphi M$ and maximal gap $\Delta_Z$, and we will see in later discussions that this seems to be indeed a distinguished point in our numerical plots. However, more work will be needed to see whether this is indeed the case.

\subsubsection{\texorpdfstring{$Z$}{Z}-channel dimension bounds}
\label{sec:Z-mixed-dimension-bounds}

\begin{figure}[htb]
\centering
\input{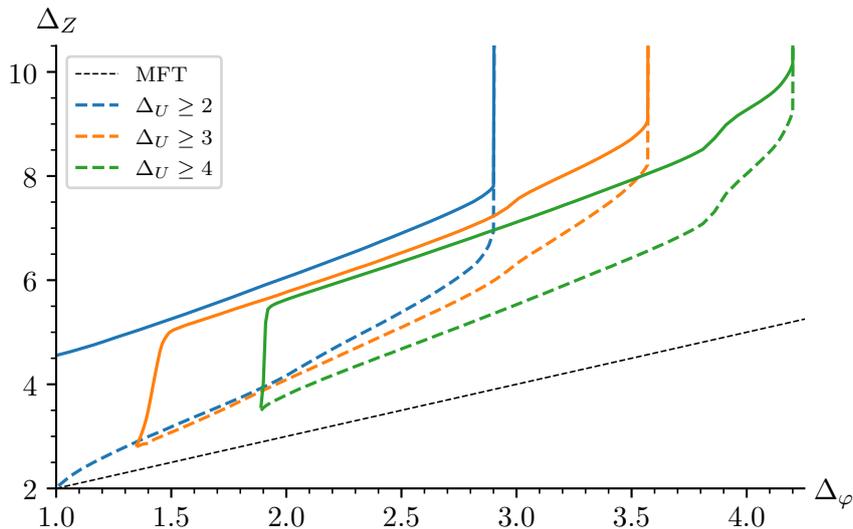}
\caption{Upper bounds on the dimension of the first unprotected multiplet in the $\varphi \times M$ OPE, as a function of the dimension of the Coulomb branch operator $\varphi$ and the gaps in the $U$ channel $\Delta_U \ge \{2, 3, 4\}$.
The solid line corresponds to the most general case, when the short operator $\varphi M$ sits at the unitarity bound $\Delta = \Delta_\varphi + 1$ and $\Delta_Z$ is the position of the first long operator.
The dashed line is obtained by the further assumption that the short operator is not present.
All results are obtained with $\Lambda = 24$.}
\label{fig:deltaPM_deltaP_deltaPPb_noCK}
\end{figure}

Let us conclude the study of conformal dimensions by putting an upper bound on the dimension $\Delta_Z$ of the first unprotected operator in the $\varphi \times M$ OPE with $\ell = 0$.
As discussed in the previous section, the most general situation is to have a protected operator $\varphi M$ at the unitarity bound $\Delta = \Delta_\varphi + 1$, after which  there is a gap until the first unprotected operator at $\Delta_Z$.
In some cases however, like the $H_1$ theory, the short operator might be missing.
One can also put gaps on the $U$ channel on top of the $Z$-channel gaps, starting with the agnostic case $\Delta_U \ge 2$.
In the numerical bounds of figure~\ref{fig:deltaPM_deltaP_deltaPPb_noCK} we consider several possibilities.
We have also explored different central charges $(c, k)$, but the results were not significantly different and therefore we assume general values for them.

The first thing that one observes in figure~\ref{fig:deltaPM_deltaP_deltaPPb_noCK} is that discarding the short operator $\varphi M$ leads to much stronger bounds than in the general case.
For theories like $H_1$, where the short is indeed missing, we have
\begin{align}
 H_1 \quad : \quad \frac{7}{3} \approx 2.33 \le \Delta_Z \le 2.90 \, .
\end{align}
Note that in figure~\ref{fig:mixed_2gap_gapZ_H1_cont} we obtained the stronger bound $2.33 \le \Delta_Z \le 2.73$, but we assumed there are only two operators in the $U$ channel with $\Delta_U, \Delta_U' \le 6$.

It is also interesting to add a gap in the $U$ channel.
From figures~\ref{fig:allowedDimPPb_deltaP_noK} and~\ref{fig:allowedDimPPb_deltaP_noC}, this restricts the dimension of the Coulomb branch operator, namely $\Delta_U \ge 3$ requires that $\Delta_\varphi \ge 1.34$ and $\Delta_U \ge 4$ requires that $\Delta_\varphi \ge 1.86$.
Indeed, when in figure~\ref{fig:mixed_2gap_gapZ_H1_cont} we assume gaps in $\Delta_U$ the curves start at the values of $\Delta_\varphi$ just discussed.
More importantly, at these values the upper bounds with and without short operator $\varphi M$ coincide.
Since there is a unique solution to crossing at the boundary of the allowed region, the solutions that saturate the bound for arbitrary $c$ and $k$ of figure~\ref{fig:allowedDimPPb_deltaP_noK} do not have the short operator $\varphi M$ in the spectrum.
This suggests that the $H_1$ theory saturates the bounds in figure~\ref{fig:allowedDimPPb_deltaP_noK}.

Finally, the reader can see from figure~\ref{fig:mixed_2gap_gapZ_H1_cont} that for large values of $\Delta_\varphi$, the upper bound of $\Delta_Z$ diverges, i.e. for $\Delta_\varphi$ high enough any value of $\Delta_Z$ is allowed.
Increasing the gap in $\Delta_U$ moves the position of this divergence to the right, but does not remove it.
We have tried different parameters of the numerical solver, either increasing the number of spins kept $\ell_{\text{max}}$, increasing the precision used by \texttt{sdpb}, or adding more poles to improve the polynomial approximation, but none of these measures has changed the results.
We currently do not know the reason of this divergence. 
It has been observed previously that mixed correlator bootstrap problems involving large external dimensions can be numerically unstable,\footnote{A possibly related issue appeared in figure 6 of~\cite{Li:2017ddj}. Instabilities in the context of mixed correlator bootstrap were discussed in~\cite{SuTalk}.} and perhaps we found another instance of these instabilities.

\subsubsection{OPE bounds}

Let us conclude our numerical exploration of the crossing equations~\eqref{eq:crossing-full} by obtaining upper and lower bounds on OPE coefficients.
As mentioned previously, we will not attempt to bound the central charges $c$ and $k$, because the analytic bounds obtained in~\cite{Liendo:2015ofa,Lemos:2015orc} are quite strong.
We will focus our efforts on studying the operators that appear in the new channel $\varphi \times M$.

\begin{figure}[htb]
\centering
\input{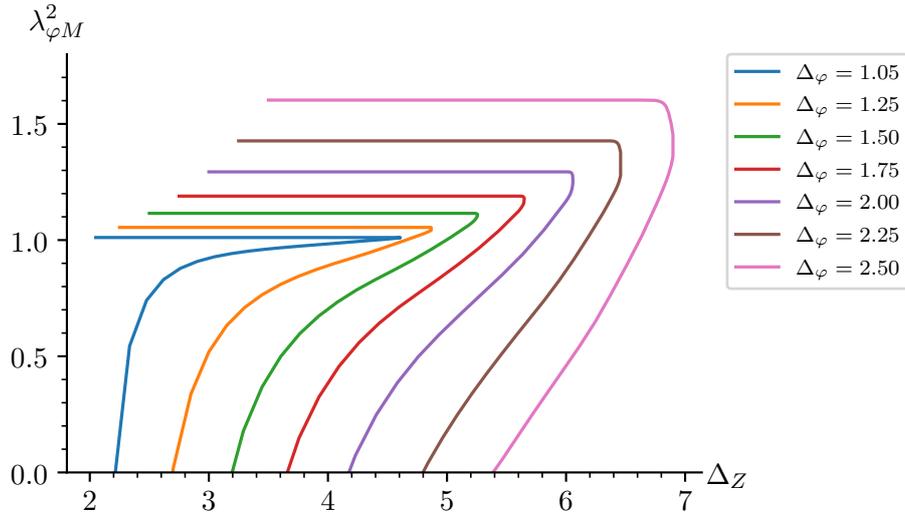}
\caption{Upper and lower bounds of the protected operator $\varphi M$ that appears in the $\varphi \times M$ OPE, as a function of the dimension $\Delta_Z$ of the first long in the same channel. 
We compare the results for different dimensions of the Coulomb branch operator $\Delta_\varphi$.
All numerical optimizations are performed at $\Lambda = 24$.}
\label{fig:opeBoundLB00_gapZ_noCK}
\end{figure}

As already discussed, restricting to scalars in the $\varphi \times M$ OPE means we have two types of operators, a short protected one at the unitarity bound that we call $\varphi M$, followed by an unprotected multiplet whose primary has dimension $\Delta_Z  > \Delta_\varphi + 1$.
In figure~\ref{fig:opeBoundLB00_gapZ_noCK}, we put upper and lower bounds on the OPE coefficient $\lambda_{\varphi,M,\varphi M}^2 \equiv \lambda_{\varphi M}^2$ as a function of the dimension of the long $\Delta_Z$.
These bounds depend strongly on the dimension of the Coulomb branch generator $\Delta_\varphi$; we have also experimented changing the central charges $c$ and $k$ but they had little influence on the final result.
When the dimension of the long is close to the unitarity bound, only upper bounds are obtained, but as we increase $\Delta_Z$ we can eventually also obtain lower bounds on the OPE coefficients.
The value of $\Delta_Z$ where the lower bound appears corresponds precisely to the maximum gap for a theory without the $\varphi M$ multiplet, namely the blue dashed line in figure~\ref{fig:deltaPM_deltaP_deltaPPb_noCK}.
As we increase $\Delta_Z$ even more, the upper and lower bounds eventually meet, and after that point there are no solutions of crossing; this corresponds to the blue solid line in figure~\ref{fig:deltaPM_deltaP_deltaPPb_noCK}.

\begin{figure}[htb]
\centering
\input{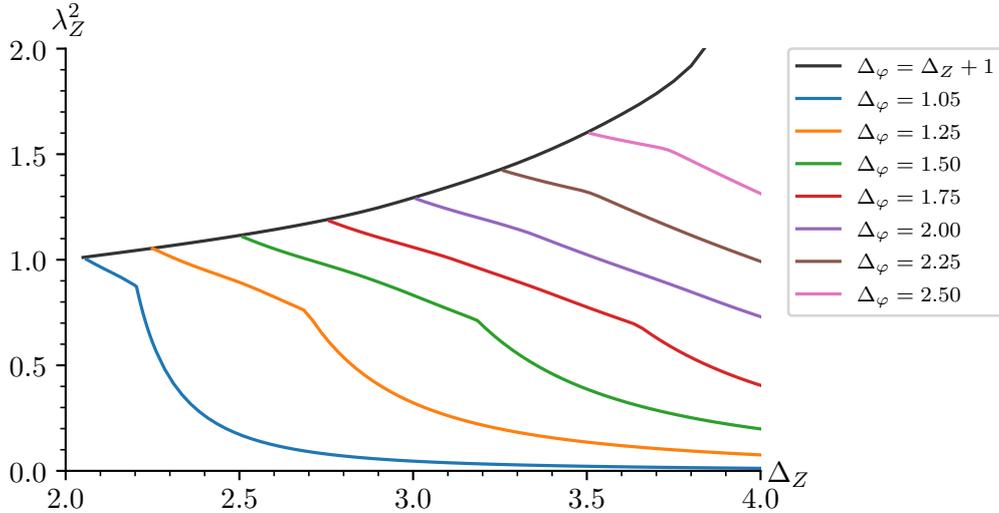}
\caption{Upper bounds on the OPE coefficient of a long scalar multiplet of dimension $\Delta_Z$ in the $\varphi \times M$ OPE, for different dimensions of the Coulomb branch operator $\varphi$.
Notice we do not assume that $\Delta_Z$ is the lowest dimension of an operator in the $Z$ channel.
The black line corresponds to the upper bound for the operator $\varphi M$ at the unitarity bound $\Delta_Z = \Delta_\varphi + 1$.
All results are obtained with $\Lambda = 24$.}
\label{fig:opeBoundLL00}
\end{figure}

One might also be interested in the OPE coefficients of the scalar long operators in the $\varphi \times M$ OPE, which are presented in figure~\ref{fig:opeBoundLL00}.
In this case we are only able to obtain upper bounds because the operators belong to a continuum.
We see that the value of these OPE coefficients decreases monotonically with the dimension $\Delta_Z$ of the operator.
At a certain value of $\Delta_Z$, which corresponds precisely to the point where $\lambda_{\varphi M}^2$ acquires a lower bound in figure~\ref{fig:opeBoundLB00_gapZ_noCK},  one can observe a kink.
From the discussion around figure~\ref{fig:deltaPM_deltaP_deltaPPb_noCK} this also seems to be the point that saturates the upper bounds of figures~\ref{fig:allowedDimPPb_deltaP_noK} and~\ref{fig:allowedDimPPb_deltaP_noC}.
Once again, it is very tempting to conjecture that the rank-one $H_1$ Argyres-Douglas theory lives at this kink.


\section{Conclusions}
\label{sec:conclusions}

In this work we appplied the numerical bootstrap to mixed correlators in $\Nm=2$ superconformal theories. We refined single-correlator bounds for OPE coefficients and compared them successfully to recent results obtained in the limit of large $r$ charge~\cite{Hellerman:2018xpi,Grassi:2019txd}. The improved bounds allowed us to extract an approximate spectrum for selected Argyres-Douglas models, information that we then used to obtain numerical islands in the space of conformal dimensions. 

We then proceeded to study a mixed system between Coulomb branch operators and the moment map. Even though these type of correlators had been analyzed before \cite{Beem:2014zpa,Lemos:2015awa,Cornagliotto:2017snu}, this is the first time mixed correlators for different types of multiplets are considered in the $4d$ $\Nm=2$ bootstrap.

As a necessary step towards the mixed system we calculated new superconformal blocks, a result interesting on its own. Recent progress that attempts to systematize the study of superblocks include the connection to Calogero-Sutherland models \cite{Buric:2019rms,Buric:2020buk}, and the analytic superspace approach of \cite{Doobary:2015gia}. Here we have contributed a new entry to the list of known superblocks, and the simplicity of our expressions hints at a simple description within the framework of \cite{Buric:2019rms,Buric:2020buk,Doobary:2015gia}.

With the mixed system at hand we applied the numerical bootstrap machinery. Figures~\ref{fig:allowedDimPPb_deltaP_noK} and~\ref{fig:allowedDimPPb_deltaP_noC} are good examples that capture how the full system behaves. As explained in the main text, there are two regimes dominated by the individual single-correlator bounds, and the mixed system smoothly interpolates the two. The fact the interpolation is smooth is slightly disappointing, because this happens in a region where we know Argyres-Douglas models live, and there is no sharp feature signalizing their presence. The bounds are nevertheless valid and give rigorous constraints on the spectrum of these theories. The mixed system also allowed us put constraints on a region of the landscape that was unexplored until now.

One possible future direction is to continue adding information in the form of extra operators. 
For the $H_0$ theory, any additional new type of multiplet will not be of the half-BPS type considered in this work, instead it will have to be semi-short or long multiplets of the superconformal algebra, which means the four-point  kinematics will be challenging. 
However, such studies can shed light on the validity of the numerical islands found in section~\ref{sec:numerics}.
Within the realm of Coulomb and Higgs branch operators, the next natural system to consider is $\varphi_{r_1}$, $\varphi_{r_2}$ and $M$. The rank-two $(A_1,A_5)$ theory\footnote{In this notation, the rank-one $H_0$, $H_1$ and $H_2$ models are $(A_1,A_2)$, $(A_1,A_3)$ and $(A_2, A_2)$ respectively.} contains precisely these type of multiplets with a $U(1)$ flavor symmetry, and although computationally this system will be more intensive than the one considered here, it is still within reach. 

A long-term direction is the inclusion of the stress-tensor multiplet. The blocks for its superconformal primary operator were obtained in \cite{Li:2018mdl}, however the numerical bootstrap has not been implemented yet. Perhaps one of the reasons this has not been done is that these this correlator is known to have nilpotent invariants \cite{Liendo:2015ofa}, which means that to impose the full $\Nm=2$ constraints one needs to include correlators of superconformal descendants. Adding the stress tensor multiplet to the correlators studied here will allow us to bootstrap all three canonical multiplets in one mixed system, and impose a huge amount of input coming from supersymmetry. Five years ago such a problem would have been clearly unfeasible, but thanks to the impressive progress on the numerical front \cite{Chester:2019ifh}, the bootstrap for correlators with several external operators is now a reality.

\acknowledgments
We thank I.~Buric, M.~Lemos, L.~Rastelli, J.~Rong, V.~Schomerus, J.~Song, and P. van Vliet for useful discussions, and D.~Orlando for interesting comments on the first version of this work.
We also thank the Simons Collaboration on the Non-perturbative Bootstrap for many stimulating activities, and  
Walter Landry in particular for installing the latest version of \texttt{sdpb} in the DESY Maxwell cluster. 
This work is supprted by the DFG through the Emmy Noether research group ``The Conformal Bootstrap Program'' project number 400570283.   

\appendix

\section{Blocks in the mixed channel}
\label{sec:app_blocks}

In this appendix we compute the superconformal blocks involving mixed chiral and moment map operators.
The arguments rely mostly on representation theory, and no calculations of three-point functions or Casimir equations are required.

\subsection{Decomposition of \texorpdfstring{$\Nm = 2$}{N=2} into \texorpdfstring{$\Nm = 1$}{N=1}}

The $\Nm = 1$ and $\Nm = 2$ superconformal algebras share the same conformal algebra, but differ in the fermionic generators.
The $\Nm = 2$ Poincare supercharges are $Q^I_\a$, $\Qb_{I\ad}$ for $I = 1, 2$ and $\a,\ad=1,2$.
We can embed the $\Nm = 1$ subalgebra by keeping only the $I = 1$ component
\begin{align}
 (Q_\alpha)_{\Nm=1} = (Q^1_\alpha)_{\Nm=2}, \qquad
 (\bar Q_{\dot \alpha})_{\Nm=1} = (\bar Q_{1 \dot \alpha})_{\Nm=2},
\end{align}
and similarly for the conformal supercharges $S_\a$ and $\Sb_\ad$. 
The $R$-symmetry is reduced from $SU(2)_R \times U(1)_r$ to $U(1)_r$ as follows
\begin{align}
\label{eq:r-charge-n1-n2}
 r_{\Nm = 1} 
 = \frac13 \left( r - 2 R_3 \right)_{\Nm = 2},
\end{align}
and each $SU(2)_R$ representation $[R]$ decomposes into the eigenvalues $R_3 = -R, -R+2, \ldots, R-2, R$. 
Now we study the implications of this decomposition for the chiral and moment map multiplets.

Remember that a chiral operator is a scalar killed by all $\Qb_{I\ad}$ supercharges, so it can also be understood as an $\Nm = 1$ chiral $\varphi_r \to \phi_{\frac{r}{3}}$, where the $r$ charge assignment follows from~\eqref{eq:r-charge-n1-n2}.
Here and in what follows, we denote the superprimary operator of an $\Nm = 1$ chiral multiplet by $\phi_r$, satisfying the shortening $[\Qb_\ad, \phi_r(0)] = 0$. The $r$ denotes its $U(1)_r$ charge, which is related to the conformal dimension by $\Delta_\phi = \frac{3}{2} r$.

On the other hand, the moment map operator is a scalar satisfying the shortening conditions~\eqref{eq:short-M}.
We can reduce them to $\Nm = 1$ shortening conditions, noting $Q_{2\a} \to -Q_\a$ and $\Qb_{1\ad} \to \Qb_\ad$:
\begin{align}
 [\Qb_\ad, M_{11}(0)] = 0, \qquad
 [Q^2, M_{12}(0)] = [\Qb^2, M_{12}(0)] = 0, \qquad
 [Q_\a, M_{22}(0)] = 0.
\end{align}
Thus the components $M_{11}$, $M_{12}$ and $M_{22}$ are $\Nm = 1$ chiral, $\Nm = 1$ current and $\Nm = 1$ antichiral operators:
\begin{align}
 M_{11} \to \phi_{\frac{4}{3}}, \qquad
 M_{12} \to J, \qquad
 M_{22} \to \bar \phi_{-\frac{4}{3}}.
\end{align}

Finally, we will use repeatedly the decomposition of $\Nm = 2$ long multiplets without $SU(2)_R$ charge into $\Nm = 1$ multiplets:
\begin{align}
\begin{split}
 L \bar L [j, \jbar]^{(0;r)}_\Delta
\to & \;
   L \bar L [j, \jbar]^{(\frac13r)}_\Delta \\
&+ \sum_{s=\pm1} L \bar L [j + s, \jbar]^{(\frac13(r+1))}_{\Delta+\frac12}
 + \sum_{\bar s=\pm1} L \bar L [j, \jbar + \bar s]^{(\frac13(r-1))}_{\Delta+\frac12} \\
&+ \sum_{s,\bar s=\pm1} L \bar L [j + s, \jbar + \bar s]^{(\frac13 r)}_{\Delta+1}
 + \sum_{s=\pm1} L \bar L [j, \jbar]^{(\frac13 (r+2s))}_{\Delta+1} \\
&+ \sum_{s=\pm1} L \bar L [j + s, \jbar]^{(\frac13(r-1))}_{\Delta+\frac32}
 + \sum_{\bar s=\pm1} L \bar L [j, \jbar + \bar s]^{(\frac13(r+1))}_{\Delta+\frac32} \\
&+ L \bar L [j, \jbar]^{(\frac13 r)}_{\Delta+2}.
\end{split}
\end{align}
This decomposition can be obtained combining the tables in~\cite{Cordova:2016emh} with the rule~\eqref{eq:r-charge-n1-n2}.

\subsection{\texorpdfstring{$\langle \varphi\bar\varphi MM\rangle$}{<phi phib M M>} correlator}

In order to compute the blocks it is convenient to switch back to component notation for the moment map.
In this section we suppress the adjoint flavor indices to simplify the notation.
We start with the four point function~\eqref{eq:corr-PPbMM}:
\begin{align}
\label{eq:corr-PPbMM-comp}
 \langle
        \varphi (x_1)
   \bar \varphi (x_2)
   M_{IJ}(x_3)
   M_{KL}(x_4)
 \rangle
 = \frac{\epsilon_{I(K} \epsilon_{J|L)}}{|x_{12}|^{2\Delta_\varphi} |x_{34}|^4} 
   \sum_{\Om \in A^+} \lambda_{\varphi\bar\varphi\Om} \lambda_{MM\Om}G^{\varphi\bar\varphi;MM}_{\Delta,\ell}(z,\bar z).
\end{align}
The sum runs over all even spin supermultiplets that appear in the chiral-antichiral and moment map OPEs:
\begin{align}
\label{eq:sel-rul-ppMM-ap}
 \varphi \times \bar \varphi, \; M \times M
 \; \sim \; 
       \mathds{1} 
   + A \bar A[\ell; \ell]^{(0; 0)}
   + L \bar L[\ell; \ell]^{(0; 0)}_\Delta, \qquad
 \ell \text{ even}.
\end{align}

By looking at the $(I,J) = (1, 1)$ and $(K,L) = (2,2)$ components, we can think of \eqref{eq:corr-PPbMM-comp} as a correlator of different $\Nm=1$ chiral and antichiral operators $\langle \phi_1 \bar \phi_1 \phi_2 \bar \phi_2 \rangle$. 
The selection rules and superconformal blocks for this correlator are well understood~\cite{Poland:2010wg,Fitzpatrick:2014oza,Khandker:2014mpa}.
The $\Nm = 1$ OPE is
\begin{align}
\label{eq:sel-rule-N1}
 & \phi_{r_i} \times \bar \phi_{-r_i}
   \sim \mathds{1} 
   + A \bar A[\ell; \ell]^{(0)}
   + L \bar L[\ell; \ell]^{(0)}_\Delta,
\end{align}
and the contribution of the long multiplet is captured by the following superconformal block:
\begin{align}
\begin{split}
\label{eq:n1-block-p1p1bp2p2b}
 G^{\phi_1\bar\phi_1,\phi_2\bar\phi_2}_{\Delta,\ell}
 = & \; g_{\Delta,\ell}
 + \frac{(\Delta + \ell)}{4(\Delta + \ell + 1)} \, g_{\Delta+1,\ell+1}
 + \frac{(\Delta - \ell - 2)}{4(\Delta - \ell - 1)} \, g_{\Delta+1,\ell-1} \\
 & + \frac{(\Delta + \ell)(\Delta - \ell - 2)}
          {16(\Delta + \ell + 1)(\Delta - \ell - 1)} \, g_{\Delta+2,\ell}.
\end{split}
\end{align}
The contribution for the exchanged $A\bar A[\ell;\ell]^{(0)}$ short multiplet is obtained setting this block at the unitarity bound $\Delta = \ell + 2$.
Notice that the $\Nm = 2$ long exchanged in~\eqref{eq:sel-rul-ppMM-ap} decomposes in $\Nm = 1$ language as
\begin{align}
   L \bar L[\ell; \ell]^{(0;0)}_\Delta
 \to
   L \bar L[\ell; \ell]^{(0)}_\Delta
 + L \bar L[\ell+1; \ell+1]^{(0)}_{\Delta+1}
 + L \bar L[\ell-1; \ell-1]^{(0)}_{\Delta+1}
 + L \bar L[\ell; \ell]^{(0)}_{\Delta+2},
\end{align}
where we have omitted all $\Nm = 1$ multiplets that are not allowed by the selection rule~\eqref{eq:sel-rule-N1}.
Thus, we must have the following decomposition:
\begin{align}
\begin{split}
\label{eq:PPbMM-block-ansatz}
 G^{\varphi\bar\varphi;MM}_{\Delta,\ell}
 & =
   a_0 G^{\phi_1\bar\phi_1;\phi_2\bar\phi_2}_{\Delta,\ell}
 + a_1 G^{\phi_1\bar\phi_1;\phi_2\bar\phi_2}_{\Delta+1,\ell+1}
 + a_2 G^{\phi_1\bar\phi_1;\phi_2\bar\phi_2}_{\Delta+1,\ell-1}
 + a_3 G^{\phi_1\bar\phi_1;\phi_2\bar\phi_2}_{\Delta+2,\ell}.
\end{split}
\end{align}
We are going to fix the coefficients $a_i$ in two different ways leading to the same result.
First, in the OPE of two moment maps only operators with even spin can appear.
Inserting~\eqref{eq:n1-block-p1p1bp2p2b} in~\eqref{eq:PPbMM-block-ansatz} and demanding that only even spin descendants contribute fixes the block completely:
\begin{align}
\label{eq:block-ppbMM-ap}
 G^{\varphi\bar\varphi;MM}_{\Delta,\ell}
 = & \; g_{\Delta ,\ell }
  - \frac{(\Delta-\ell) (\Delta - \ell -2)}{16 (\Delta - \ell + 1) (\Delta - \ell -1)}
    g_{\Delta +2,\ell -2}
  - \frac{(\Delta +\ell ) (\Delta +\ell +2)}{16 (\Delta +\ell +1) (\Delta +\ell +3)}
    g_{\Delta +2,\ell +2} \nonumber \\
 & + \frac{(\Delta-\ell) (\Delta - \ell -2)(\Delta +\ell ) (\Delta +\ell +2)}
     {256 (\Delta - \ell + 1) (\Delta - \ell -1) (\Delta +\ell +1) (\Delta +\ell +3)} 
     g_{\Delta +4,\ell }.
\end{align}
As discussed, this block corresponds to the exchange of a generic long multiplet.
The contribution of $A\bar A[\ell;\ell]^{(0;0)}$ is obtained by evaluating the block at the unitarity bound $\Delta = \ell + 2$. 
By using hypergeometric identities one can check that~\eqref{eq:block-ppbMM-ap} is identical to the compact expression provided in the main text~\eqref{eq:n2-block-ppbMM}.

Let us derive the same result in an alternative way.
Looking at the $(I,J)=(K,L)=(1,2)$ component of~\eqref{eq:corr-PPbMM-comp} we get a correlation function of $\Nm = 1$ chirals and  currents $\langle \phi \bar \phi J J \rangle$.
As before, the selection rules and superconformal blocks are well understood~\cite{Fortin:2011nq,Khandker:2014mpa,Berkooz:2014yda}.
In particular, the only multiplet relevant in the OPE is $J \times J
   \sim L\bar L[\ell; \ell]^{(0)}_\Delta$, for which the conformal block is
\begin{align}
 G^{\phi\bar\phi,JJ}_{\Delta,\ell \text{ even}}
 & = g_{\Delta ,\ell }
   - \frac{(\Delta -2) (\Delta -\ell -2) (\Delta +\ell )}
          {16 \Delta  (\Delta -\ell -1) (\Delta +\ell +1)}
          g_{\Delta +2,\ell }, \\
 G^{\phi\bar\phi,JJ}_{\Delta,\ell \text{ odd}}
 & = g_{\Delta +1,\ell +1}
   - \frac{(\ell +2) (\Delta -\ell -2) (\Delta +\ell +1)}
          {\ell  (\Delta -\ell -1) (\Delta +\ell )}
          g_{\Delta +1,\ell -1}.
\end{align}
Following the same reasoning as before, the $\Nm = 2$ block should decompose as
\begin{align}
\begin{split}
\label{eq:PPbJJ-block-ansatz}
 G^{\varphi\bar\varphi;MM}_{\Delta,\ell}
 & = b_0 G^{\phi\bar\phi;JJ}_{\Delta,\ell,\text{even}}
   + b_1 G^{\phi\bar\phi;JJ}_{\Delta+1,\ell+1,\text{odd}}
   + b_2 G^{\phi\bar\phi;JJ}_{\Delta+1,\ell-1,\text{odd}}
   + b_3 G^{\phi\bar\phi;JJ}_{\Delta+2,\ell,\text{even}}.
\end{split}
\end{align}
Demanding that the two decompositions~\eqref{eq:PPbMM-block-ansatz} and~\eqref{eq:PPbJJ-block-ansatz} are equal uniquely determines the superconformal block~\eqref{eq:block-ppbMM-ap}.

\subsection{\texorpdfstring{$\langle \varphi MM \bar\varphi \rangle$}{<phi M M phib>} correlator}

Now we move on to the calculation of the blocks in the crossed channel:
\begin{align}
\begin{split}
 \langle
        \varphi (x_1)
 &       M_{IJ} (x_2)
         M_{KL} (x_3)
   \bar \varphi (x_4)
 \rangle \\
 & = \frac{\epsilon_{I(K} \epsilon_{J|L)}}
        {x_{12}^{\Delta_\varphi + 2} x_{34}^{\Delta_\varphi + 2}} 
   \left( \frac{x_{24}}{x_{14}} \right)^{\Delta_\varphi - 2}
   \left( \frac{x_{14}}{x_{13}} \right)^{\Delta_\varphi - 2}
   \sum_\Om |\lambda_{\varphi M \Om}|^2 \, 
            G^{\varphi M;M\bar\varphi}_{\Delta,\ell}(z,\bar z) \, .
\end{split}
\end{align}
The first important question is what multiplets appear in the sum over superdescendants.
Using the superspace calculation of~\cite{Ramirez:2016lyk}, we obtain the following selection rule:
\begin{align}
\label{eq:sel-rul-pM-ap}
 \varphi_r \times M
 \sim
    L \bar B[0; 0]^{(2;r)}
  + L \bar A[\ell; \ell-1]^{(1;r-1)}
  + L \bar L[\ell; \ell]_\Delta^{(0;r-2)} \, .
\end{align}

The strategy to compute the conformal blocks is the same as before.
Consider the component $(I,J) = (2,2)$ and $(K, L) = (1, 1)$ of the four-point function and interpret it as a correlator of $\Nm = 1$ chirals and antichirals $\langle \phi_1 \bar \phi_2 \phi_2 \bar \phi_1 \rangle$.
The selection rules and blocks are once again in the literature~\cite{Lemos:2015awa}.
Restricting to the case of interest to us, we have
\begin{align}
\label{eq:sel-rul-ppb-n1-ap}
 \phi_{\frac r3} \times \bar \phi_{\text-\frac43}
 \sim
   L\bar B[0; 0]^{(\frac{r\text-4}{3})}
 + L\bar A[\ell; \ell]^{(\frac{r\text-4}{3})}
 + L\bar L[\ell; \ell]^{(\frac{r\text-4}{3})}_\Delta \, ,
\end{align}
and the contribution of the long is given by the superconformal block:
\begin{align}
 G^{\phi_1\bar \phi_2; \phi_2\bar \phi_1}_{\Delta,\ell}(z,\bar z)
 = (z \bar z)^{-1/2} g_{\Delta+1,\ell}^{1+\Delta_{12},1-\Delta_{12}}(z,\bar z) \, .
\end{align}
As before, we want to write our $\Nm = 2$ block as a linear combination of $\Nm = 1$ ones. For concreteness, let us focus on the exchange of the long in~\eqref{eq:sel-rul-pM-ap}, which has the following decomposition:
\begin{align}
\begin{alignedat}{2}
& L \bar L[\ell; \ell]_\Delta^{(0;r\text-2)}
&& \to \;
  L\bar L[\ell; \ell]^{(\frac{r\text-4}{3})}_{\Delta+1} \, .
\end{alignedat}
\end{align}
We have dropped all terms that are not compatible with the $\Nm = 1$ selection rule~\eqref{eq:sel-rul-ppb-n1-ap}.
Since there is only one $\Nm=1$ multiplet in the decomposition, the block must be equal to its $\Nm = 1$ counterpart:
\begin{align}
 G^{\varphi M;M\bar \varphi}_{\Delta,\ell}(z,\zb)
 = G^{\phi_1 \bar\phi_2;\phi_2 \bar\phi_1}
    _{\Delta+1,\ell}(z,\zb)
 = (z \zb)^{-1/2} g_{\Delta+2,\ell}^{\Delta_\varphi - 1, 3 - \Delta_\varphi}(z,\zb) \,.
\end{align}
Naturally, this can be expanded into non-supersymmetric blocks
\begin{align}
\label{eq:decomposition-pMMpb-ap}
 G^{\varphi M;M\bar \varphi}_{\Delta,\ell} 
 =         g^{\Delta_\varphi - 2, 2-\Delta_\varphi}_{\Delta+1, \ell}
 + c_1     g^{\Delta_\varphi - 2, 2-\Delta_\varphi}_{\Delta+2, \ell+1}
 + c_2     g^{\Delta_\varphi - 2, 2-\Delta_\varphi}_{\Delta+2, \ell-1}
 + c_1 c_2 g^{\Delta_\varphi - 2, 2-\Delta_\varphi}_{\Delta+3, \ell},
\end{align}
with
\begin{align}
\begin{split}
 c_1 & = \frac{(\Delta -\Delta_\varphi+\ell+3) (\Delta + \Delta_\varphi+\ell-1)}
              {4 (\Delta +\ell+1) (\Delta +\ell+2)}, \\[0.5em]
 c_2 & = \frac{(\Delta - \Delta_\varphi - \ell + 1) (\Delta + \Delta_\varphi -\ell-3)}
              {4 (\Delta -\ell-1) (\Delta -\ell)}.
\end{split}
\end{align}

As a non-trivial sanity check, we can instead look at the $(I,J)=(K,L)=(1,2)$ component of the four-point function, and interpret it as a $\Nm = 1$ correlator $\langle \phi J J \phi \rangle$.
The selection rule in this case is~\cite{Li:2017ddj}
\begin{align}
\label{eq:ope-pJ-n1-ap}
 \phi_{r} \times J
 \sim
   L\bar L[\ell; \ell + 1]^{(r-1)}_\Delta
 + L\bar L[\ell; \ell - 1]^{(r-1)}_\Delta
 + L\bar L[\ell; \ell]^{(r-2)}_\Delta
 + \text{shorts}.
\end{align}
Only the first two will play a role, and the associated blocks are
\begin{align}
\begin{split}
 L\bar L[\ell; \ell + 1]^{(r-1)}_\Delta: \quad
   \hat G_{\Delta,\ell} =
   \hat c_1 \, g_{\Delta + 1/2,\ell  }^{\Delta_\phi-2,2-\Delta_{\phi}}
 + \hat c_2 \, g_{\Delta + 3/2,\ell+1}^{\Delta_\phi-2,2-\Delta_{\phi}}, \\[0.5em]
 L\bar L[\ell; \ell - 1]^{(r-1)}_\Delta: \quad
   \check G_{\Delta,\ell} =
   \check c_1 \, g_{\Delta + 1/2,\ell  }^{\Delta_\phi-2,2-\Delta_{\phi}}
 + \check c_2 \, g_{\Delta + 3/2,\ell-1}^{\Delta_\phi-2,2-\Delta_{\phi}},
\end{split}
\end{align}
with coefficients~\cite{Li:2017ddj}:
\begin{align}
\begin{split}
 \hat c_1 
 & = \frac{(\ell +2)}
          {(\ell +1) (2 \Delta - 2 \Delta _{\phi } - 2 \ell - 3)}, \\[0.5em]
 \hat c_2 
 & = \frac{(2 \Delta -3) (2 \Delta -2 \Delta _{\phi } + 2 \ell +5 ) 
           (2 \Delta + 2 \Delta _{\phi } +2 \ell -3) }
          {4 (2 \Delta -1) (2 \Delta +2 \ell +1) (2 \Delta +2 \ell +3) 
           (2 \Delta-2 \Delta _{\phi } -2 \ell -3 )}, \\[0.5em]
 \check c_1 
 & = \frac{1}{(2 \Delta -2 \Delta _{\phi } +2 \ell +1)}, \\[0.5em]
 \check c_2 
 & = \frac{(2 \Delta -3) (\ell +1) (2 \Delta - 2 \Delta _{\phi } -2 \ell +1)
           (2 \Delta + 2 \Delta _{\phi } - 2 \ell - 7 )}
          {4 (2 \Delta -1) \ell  (2 \Delta -2 \ell -1) (2 \Delta -2 \ell -3)
           (2 \Delta -2 \Delta _{\phi } +2 \ell +1 )}.
\end{split}
\end{align}
Decomposing the $\Nm = 2$ long multiplet and keeping only terms compatible with the OPE~\eqref{eq:ope-pJ-n1-ap} we get
\begin{align}
\begin{split}
 L \bar L[\ell; \ell]_\Delta^{(0;r\text-2)}
 \to & \;
    L\bar L[\ell; \ell+1]^{(\frac{r\text-3}{3})}_{\Delta+\frac12}
  + L\bar L[\ell; \ell-1]^{(\frac{r\text-3}{3})}_{\Delta+\frac12} \\
& + L\bar L[\ell+1; \ell]^{(\frac{r\text-3}{3})}_{\Delta+\frac32}
  + L\bar L[\ell-1; \ell]^{(\frac{r\text-3}{3})}_{\Delta+\frac32} \, ,
\end{split}
\end{align}
so we find the following decomposition:
\begin{align}
\label{eq:PMMPb-decomposition-JJ}
 G^{\varphi M;M\bar \varphi}_{\Delta,\ell}
 = d_0 \, \hat   G^{\phi J; J \bar\phi}_{\Delta+1/2,\ell}
 + d_1 \, \check G^{\phi J; J \bar\phi}_{\Delta+1/2,\ell}
 + d_2 \, \check G^{\phi J; J \bar\phi}_{\Delta+3/2,\ell+1}
 + d_3 \, \hat   G^{\phi J; J \bar\phi}_{\Delta+3/2,\ell-1} \, .
\end{align}
There is a linear relation between the four $\Nm = 1$ blocks above, and we fix it by setting $d_0 = 0$.
It is an easy exercise to check that using the remaining coefficients we can indeed obtain the decomposition~\eqref{eq:PMMPb-decomposition-JJ}:
\begin{align}
\begin{split}
 & d_1 = 2 (\Delta - \Delta_\varphi + \ell + 1), \\
 & d_2 = \frac{(\Delta + \Delta_\varphi +\ell-1)(\Delta -\Delta_\varphi + \ell +3)^2}{2 (\Delta +\ell+1) (\Delta +\ell+2)}, \\
 & d_3 = -\frac{(\Delta + \Delta_\varphi  - \ell - 3) (\Delta - \Delta_\varphi - \ell + 1)^2}
              {2 \Delta  (\ell+1) (\Delta - \ell)}.
\end{split}
\end{align}
The reader can check that any minor change to the $\Nm = 2$ or $\Nm = 1$ superblocks prevents this decomposition from being possible.
This provides a very non-trivial check for our superblock~\eqref{eq:decomposition-pMMpb-ap}, as well as for the results in~\cite{Li:2017ddj}.

\section{Numerical implementation}
\label{sec:app_numerics}

\subsection{Approximating blocks by polynomials}
\label{sec:app_numerics-blocks}

In order to build polynomial approximations of the four-dimensional conformal blocks we use the recursion relations originally obtained in \cite{Poland:2010wg,Poland:2011ey}, and later generalized in~\cite{Li:2017ddj}.

The quantities we need to approximate are derivatives of one-dimensional conformal blocks, evaluated at the crossing-symmetric point:
\begin{align}
 C_{\alpha,\beta,\gamma,\delta}^n 
 = \frac{\partial^n }{\partial z^n} 
   \left( z^{1-\delta} k_{\alpha}^{\beta,\gamma}(z) \right)_{z=1/2}.
\end{align}
We take $\beta$, $\gamma$, $\delta$ and $n$ to have fixed numerical values, and we wish to approximate $C_{\alpha,\beta,\gamma,\delta}$ as a polynomial in $\alpha$ times a positive function of $\alpha$.
Using the differential equation satisfied by $k_{\alpha}^{\beta,\gamma}(z)$, one obtains the following recursion relation~\cite{Li:2017ddj}:
\begin{align}
\begin{split}
 C_{\alpha,\beta,\gamma,\delta}^n 
 = 
& - (2n + \beta - \gamma + 4 \delta - 10) C_{\alpha,\beta,\gamma,\delta}^{n-1} \\
& + \big(
   4n (n - \beta + \gamma - 3) \\
&  \qquad \quad
   + 2 \alpha(\alpha - 2)
   - \beta ( \gamma + 2 \delta - 10)
   + \gamma ( 2 \delta - 10)
   - 4 \delta ( \delta - 4) - 4
  \big) C_{\alpha,\beta,\gamma,\delta}^{n-2} \\
& + 2 (n - 2)(2n - \beta + 2\delta-8)(2n + \gamma + 2 \delta - 8) 
   C_{\alpha,\beta,\gamma,\delta}^{n-3}.
\end{split}
\end{align}
Applying the recursion repeatedly, we can write
\begin{align}
 C_{\alpha,\beta,\gamma,\delta}^n 
 = P_n(\alpha,\beta,\gamma,\delta) \, k_{\alpha}^{\beta,\gamma}(1/2)
 + Q_n(\alpha,\beta,\gamma,\delta)
 \frac{\partial k_{\alpha}^{\beta,\gamma}}{\partial z}\left(1 / 2 \right) \, ,
\end{align}
where $P_n$ and $Q_n$ are polynomials in $\alpha$.
Next, we should approximate $k_{\alpha}^{\beta,\gamma}(1/2)$ and its first derivative as polynomials in $\alpha$ times a positive prefactor. We introduce the radial coordinate $\rho$~\cite{Hogervorst:2013sma}
\begin{align}
 z = \frac{4\rho}{(1+\rho)^2},
\end{align}
and expand the one-dimensional blocks in a power series in $\rho$ up to order $w$.
Finally, we evaluate the expansion at the crossing symmetric point $\rho_* = 3-2\sqrt{2}\approx 0.17$.
Remember that the one-dimensional blocks are schematically $z^{\alpha/2}$ times a ${}_2F_1$ hypergeometric, so the expansion will have the form
\begin{align}
 k_\alpha^{\beta,\gamma}(1/2) 
 \approx (4\rho_*)^{\alpha/2} \sum_{j=0}^w R_j(\alpha) \rho_*^j
 = \frac{(4\rho_*)^{\alpha/2}}{D_{\beta,\gamma,w}(\alpha)} N_{\beta,\gamma,w}(\alpha), \\
 \frac{\partial k_\alpha^{\beta,\gamma}}{\partial z}(1/2)
 \approx (4\rho_*)^{\alpha/2} \sum_{j=0}^{w-1} S_j(\alpha) \rho_*^j
 = \frac{(4\rho_*)^{\alpha/2}}{D_{\beta,\gamma,w}(\alpha)} L_{\beta,\gamma,w}(\alpha),
\end{align}
where the terms $R_j(\alpha), S_j(\alpha)$ are rational functions of $\alpha$, and so we can factor a common term $(4\rho_*)^{\alpha/2} / D(\alpha)$.
The zeros of $D(\alpha)$ are at $\alpha = 0, -1, -2, \ldots$, and one can see that the prefactor is always positive provided the unitary bounds are satisfied.
Combining these ingredients it is a simple exercise to construct the polynomial approximation of the conformal blocks.

There are other possible approaches to compute the polynomial approximations.
One idea would be to use the Zamolodchikov-like recursion relations for $k_{\alpha}^{\beta,\gamma}(z)$ derived in~\cite{Karateev:2019pvw}.
Another option would be to use the recursion relations directly in $4d$~\cite{Kravchuk:2019}, which are already implemented in \texttt{scalar\_blocks}.
It would be interesting to compare the different approaches in terms of performance, accuracy of the blocks and size of the polynomial approximation.

\subsection{SDPB parameters}
\label{sec:app_numerics-sdpb}

In this work, we have computed mixed correlator bounds at $\Lambda = 24$ and single correlator bounds at $\Lambda = 32, 50$ using \texttt{sdpb}~\cite{Simmons-Duffin:2015qma, Landry:2019qug}.
As we increase the number of derivatives it is necessary to increase the spins included to ensure numerical stability:
\begin{align}
\begin{split}
 S_{\Lambda=24} 
 & = \{0, \ldots, 26\} \cup \{29, 30, 33, 34, 37, 38, 41, 42, 45, 46 \}, \\
 S_{\Lambda=32} 
 & = \{0, \ldots, 44\} \cup \{47, 48, 51, 52, 55, 56, 59, 60, 63, 64, 67, 68 \}, \\
 S_{\Lambda=50} 
 & = \{0, \ldots, 64\} \cup \{67, 68, 71, 72, 75, 76, 79, 80, 83, 84, 87, 88 \}.
\end{split}
\end{align}
Furthermore, it is also important to increase the accuracy of the polynomial approximation $w$ and the precision used by \texttt{sdpb}:
\begin{align}
\begin{tabular}{c|c|c|c}
 & $\Lambda=24$ & $\Lambda=32$ & $\Lambda=50$ \\ \hline
 $w$ & $18$ & $18$ & $26$ \\
 \texttt{prec} & $768$ & $768$ & $1024$.
\end{tabular}
\end{align}
We have observed that the polynomial approximation of section~\ref{sec:app_numerics-blocks} works better for single correlators. This is the reason why for the single correlator bootstrap at $\Lambda = 32$ we can keep the same number of poles as for the mixed bootstrap at $\Lambda = 24$.
For the dimension bounds, we rely heavily on the hot-starting procedure introduced in~\cite{Go:2019lke}, which speeds up the computations significantly.
We instruct \texttt{sdpb} to stop as soon as a primal or dual feasible solutions is found:
\begin{align}
\begin{split}
 & \mathtt{findPrimalFeasible}
 = \mathtt{findDualFeasible}
 = \mathrm{true}, \\
 & \mathtt{detectPrimalFeasibleJump}
 = \mathtt{detectDualFeasibleJump}
 = \mathrm{true}.
\end{split}
\end{align}
In practice, we observed that the algorithm always stopped after a primal or dual jump.
For the OPE optimizations, in order to speed them up, we have lowered the default \texttt{dualityGapThreshold} to
\begin{align}
 \mathtt{dualityGapThreshold} = 10^{-10}.
\end{align}
For the remaining parameters, we have found that the defaults of \texttt{sdpb} lead to stable results.

\subsection{Crossing vectors}
\label{sec:cross-vectors}

In this appendix we write the explicit crossing vectors that appear in equation~\eqref{eq:crossing-full}.
The $\vec I_{c,k}$ term captures all the known contributions, either from the identity, stress-tensor or flavor current exchanges.
The normalization of the stress-tensor contribution for the mixed blocks can be obtained from~\eqref{eq:stress-tens-PPMM}.
The rest has already been discussed in the main text:
\begin{align}
 \vec I_{c,k} =
  \begin{pmatrix}[1.5]
  \ph- 4 \Hm_{+,\mathbf 1}^\text{short} 
     + 3 \Hm_{+,\mathbf 3}^\text{short} 
     + 5 \Hm_{+,\mathbf 5}^\text{short} \\
  \ph- 2 \Hm_{+,\mathbf 1}^\text{short}
     + 9 \Hm_{+,\mathbf 3}^\text{short}
     - 5 \Hm_{+,\mathbf 5}^\text{short} \\
     - 2 \Hm_{-,\mathbf 1}^\text{short}
     + 3 \Hm_{-,\mathbf 3}^\text{short}
     + 5 \Hm_{-,\mathbf 5}^\text{short} \\
  E^{\varphi\bar\varphi;\varphi\bar\varphi}_{+,0,0} 
  + \frac{\Delta_\varphi^2}{6c} E^{\varphi\bar\varphi;\varphi\bar\varphi}_{+,2,0} \\
  E^{\varphi\bar\varphi;\varphi\bar\varphi}_{+,0,0}
  + \frac{\Delta_\varphi^2}{6c} 
    \tilde E^{\varphi\bar\varphi;\varphi\bar\varphi}_{+,2,0} \\
  E^{\varphi\bar\varphi;\varphi\bar\varphi}_{-,0,0}
  + \frac{\Delta_\varphi^2}{6c} 
    \tilde E^{\varphi\bar\varphi;\varphi\bar\varphi}_{-,2,0} \\
  0                                                 \\
  E^{MM;\varphi\bar\varphi}_{+,0,0}
  - \frac{\Delta_\varphi}{6c} E^{MM;\varphi\bar\varphi}_{+,2,0} \\
  E^{MM;\varphi\bar\varphi}_{-,0,0}
  - \frac{\Delta_\varphi}{6c} E^{MM;\varphi\bar\varphi}_{-,2,0} \\
 \end{pmatrix}
 \end{align}
The remaining crossing vectors can be easily obtained as discussed in the main text:
\begin{align}
 U_{\Delta,\ell} =
 \begin{pmatrix}
  \begin{pmatrix}  4 E_{+,\Delta,\ell}^{MM,MM} & 0 \\ 0 & 0 \end{pmatrix}  \\
  \begin{pmatrix}  2 E_{+,\Delta,\ell}^{MM,MM} & 0 \\ 0 & 0 \end{pmatrix}  \\
  \begin{pmatrix} -2 E_{-,\Delta,\ell}^{MM,MM} & 0 \\ 0 & 0 \end{pmatrix}  \\
  \begin{pmatrix} 0 & 0 
  \\ 0 & E_{+,\Delta,\ell}^{\varphi\bar\varphi;\varphi\bar\varphi} \end{pmatrix}  \\
  \begin{pmatrix} 0 & 0 
  \\ 0 & \tilde E_{+,\Delta,\ell}^{\varphi\bar\varphi;\varphi\bar\varphi} \end{pmatrix}  \\
  \begin{pmatrix} 0 & 0 
  \\ 0 & \tilde E_{-,\Delta,\ell}^{\varphi\bar\varphi;\varphi\bar\varphi} \end{pmatrix}  \\
  0 \\ 
  \begin{pmatrix} 
    0 & \frac12 E_{+,\Delta,\ell}^{MM,\varphi\bar\varphi} \\ 
    \frac12 E_{+,\Delta,\ell}^{MM,\varphi\bar\varphi} & 0 \end{pmatrix}  \\
  \begin{pmatrix} 
    0 & \frac12 E_{-,\Delta,\ell}^{MM,\varphi\bar\varphi} \\ 
    \frac12 E_{-,\Delta,\ell}^{MM,\varphi\bar\varphi} & 0 \end{pmatrix}  \\
 \end{pmatrix}, \quad
 V_{\Delta,\ell} =
 \begin{pmatrix}[1.3]
  0 \\
  0 \\
  0 \\
         E_{+,\Delta,\ell}^{\varphi \bar\varphi;\varphi\bar\varphi} \\
  \tilde E_{+,\Delta,\ell}^{\varphi \bar\varphi;\varphi\bar\varphi} \\
  \tilde E_{-,\Delta,\ell}^{\varphi \bar\varphi;\varphi\bar\varphi} \\
  0 \\
  0 \\
  0
 \end{pmatrix}, \quad
 W_{\Delta,\ell} =
 \begin{pmatrix}[1.3]
  0 \\
  0 \\
  0 \\
  0 \\
  \ph- E_{+,\Delta,\ell}^{\varphi\varphi;\bar\varphi\bar\varphi} \\
     - E_{-,\Delta,\ell}^{\varphi\varphi;\bar\varphi\bar\varphi} \\
  0 \\
  0 \\
  0
 \end{pmatrix}, \qquad
\end{align}
\begin{align}
 X_{\Delta,\ell} =
 \begin{pmatrix}[1.3]
  3 \, E_{+,\Delta,\ell}^{MM,MM} \\
  9 \, E_{+,\Delta,\ell}^{MM,MM} \\
  3 \, E_{-,\Delta,\ell}^{MM,MM} \\
  0 \\
  0 \\
  0 \\
  0 \\
  0 \\
  0 \\
 \end{pmatrix}, \quad
 Y_{\Delta,\ell} =
 \begin{pmatrix}[1.3]
  \ph-5 \, E_{+,\Delta,\ell}^{MM,MM} \\
     -5 \, E_{+,\Delta,\ell}^{MM,MM} \\
  \ph-5 \, E_{-,\Delta,\ell}^{MM,MM} \\
  0 \\
  0 \\
  0 \\
  0 \\
  0 \\
  0 \\
 \end{pmatrix}, \quad
 Z_{\Delta,\ell} =
 \begin{pmatrix}[1.3]
  0 \\
  0 \\
  0 \\
  0 \\
  0 \\
  0 \\
  \ph- \tilde E_{+,\Delta,\ell}^{\varphi M;M\bar\varphi} \\
         \ph- E_{+,\Delta,\ell}^{\varphi M;M\bar\varphi}  \\
            - E_{-,\Delta,\ell}^{\varphi M;M\bar\varphi}
 \end{pmatrix}.
\end{align}

\subsection{Spectrum extraction}

In table~\ref{tab:spectrum} we present the results from extracting the spectrum that saturates the OPE bounds in table~\ref{tab:OPE-bounds}.
We use the package \texttt{spectrum-extraction} developed for~\cite{Komargodski:2016auf}, which is available online in the Bootstrap Collaboration website.
Although we have not collected the spectrum for $\ell > 0$ operators, we are happy to provide the data upon request. For a detailed analysis of the spectrum in the large-$\ell$ limit for the $H_0$ theory we refer to~\cite{Cornagliotto:2017snu}.

{
\renewcommand{\arraystretch}{1.5}
\renewcommand\tabcolsep{8pt}
\begin{table}[H]
\centering
\begin{tabular}{ | c | c | c | c | c | c | c | c | }
 \rule{1cm}{0pt} &
 \rule{1cm}{0pt} &
 \rule{1cm}{0pt} &
 \rule{1cm}{0pt} &
 \rule{1cm}{0pt} &
 \rule{1cm}{0pt} &
 \rule{1cm}{0pt} &
 \rule{1cm}{0pt} \\ [-\arraystretch\normalbaselineskip]
  \hline
  {\bf Theory}  & 
  {\bf Bound}  & 
  {\bf Type}  & 
  $\Delta_{U}$ & 
  $\Delta_{U}'$ & 
  $\Delta_{U}''$ & 
  $\Delta_{W}$ & 
  $\Delta_{W}'$ \\ \hline \hline
$H_0$ & $\varphi^2$ & upper & 2.70 & 5.94 & 9.28 & 4.82 & 7.82 \\ \hline
$H_0$ & $\varphi^2$ & lower & 2.66 & 5.82 & 9.14 & 4.95 & 7.57 \\ \hline
$H_0$ & $\bar Q^2 \mathcal{O}$ & upper & 2.69 & 5.88 & 9.16 & 4.81 & 7.79 \\ \hline
$H_0$ & $\bar Q^2 \mathcal{O}$ & lower & 2.66 & 5.79 & 9.07 & 4.69 & 6.49 \\ \hline
\hline
$H_1$ & $\varphi^2$ & upper & 3.05 & 6.12 & 13.11 & 5.28 & 8.27 \\ \hline
$H_1$ & $\varphi^2$ & lower & 2.92 & 5.79 & 12.79 & 5.79 & 7.74 \\ \hline
$H_1$ & $\bar Q^2 \mathcal{O}$ & upper & 3.03 & 5.98 & 9.12 & 5.27 & 8.28 \\ \hline
$H_1$ & $\bar Q^2 \mathcal{O}$ & lower & 2.92 & 5.76 & 12.65 & 4.78 & 6.35 \\ \hline
\hline
$H_2$ & $\varphi^2$ & upper & 3.45 & 6.41 & 9.66 & 5.88 & 8.96 \\ \hline
$H_2$ & $\varphi^2$ & lower & 3.23 & 5.91 & 8.98 & 6.88 & 9.89 \\ \hline
$H_2$ & $\bar Q^2 \mathcal{O}$ & upper & 3.42 & 6.29 & 9.49 & 5.85 & 8.95 \\ \hline
$H_2$ & $\bar Q^2 \mathcal{O}$ & lower & 3.21 & 3.75 & 5.97 & 6.86 & 9.38 \\ \hline
\end{tabular} 
\caption{Approximate spectrum from OPE bounds at $\Lambda = 50$.}
\label{tab:spectrum}
\end{table}
}


\providecommand{\href}[2]{#2}\begingroup\raggedright\endgroup

\end{document}